\documentclass[twocolumn]{aastex701}

\usepackage{array, booktabs}

\begin{document}


\title{From Anoxic to Oxic Earth: Detectability of Atmospheric Biosignatures across Stellar Hosts with ELT/ANDES}

\author[0000-0002-7033-209X]{Dwaipayan Dubey}
\affiliation{Universitäts-Sternwarte, Fakultät für Physik, Ludwig-Maximilians-Universität München, Scheinerstr. 1, D-81679 München, Germany}
\affiliation{Exzellenzcluster `Origins’, Boltzmannstr. 2, D-85748 Garching, Germany}
\email[show]{dwaipayan.dubey25@gmail.com}

\author[0000-0003-3652-3690]{Spandan Dash}
\affiliation{Université Paris Cité, Institut de physique du globe de Paris, CNRS, F-75005 Paris, France}
\email{dash@ipgp.fr}

\author[0000-0001-7031-8039]{Liton Majumdar}
\affiliation{Exoplanets and Planetary Formation Group, School of Earth and Planetary Sciences, National Institute of Science Education and Research, Jatni - 752050, Odisha, India}
\affiliation{Homi Bhabha National Institute, Training School Complex, Anushaktinagar, Mumbai 400094, India}
\email{liton@niser.ac.in; dr.liton.majumdar@gmail.com}


\begin{abstract}
Earth's atmosphere evolved from an anoxic, methane-bearing Archean state through Proterozoic oxygenation to the O$_2$-rich modern atmosphere. However, most biosignature detectability studies focus on modern Earth, leaving earlier atmospheric states comparatively unexplored. We investigate how atmospheric evolution and stellar environments influence the detectability of O$_2$, CH$_4$, H$_2$O, and CO$_2$, and their potential to distinguish anoxic from oxygenated atmospheres. We construct two complementary photochemical grids: representative Archean, Proterozoic, and modern-Earth scenarios around the Sun, and four geological epochs (3.9, 2.0, 0.8\,Ga, and modern Earth) across FGKM hosts. Combining one-dimensional photochemical modeling, high-resolution transmission spectral simulations, and instrument-noise modeling, we assess molecular detectability with ELT/ANDES under optimistic photon-noise-limited conditions. For solar-host Archean and Proterozoic models, CH$_4$ requires approximately 33 and 42 transits, respectively, to reach the adopted 5$\sigma$ criterion, whereas O$_2$ remains below 3$\sigma$ within the observing limits. Modern-Earth O$_2$ reaches 3$\sigma$ after approximately 38 transits. Across the FGKM grid, O$_2$ does not reach 5$\sigma$ for the 3.9\,Ga models. However, For the M3V, K2V, and M8V proxies, modern-Earth O$_2$ reaches 5$\sigma$ in 9, 15, and 20 transits, respectively, whereas CH$_4$ requires 11, 12, and 28 transits at 3.9\,Ga and only 2--4 transits across the oxygenated epochs. These results reveal a strong O$_2$ detectability contrast around M and K-dwarfs, but not Sun-like hosts in the YJH bands. Robust biosignature interpretation requires complementary molecular constraints, including informative non-detections, within the context of atmospheric evolution and stellar environment.
\end{abstract}

\keywords{\uat{High resolution spectroscopy}{2096}; \uat{Exoplanet atmospheres}{487}; \uat{Astrobiology}{74}; \uat{Biosignatures}{2018};  \uat{Terrestrial planets}{797}}

\section{Introduction}
\label{sec:intro}

The remote search for life beyond the Solar System is necessarily an exercise in comparative planetology with Earth. Observations of our own planet, from spacecraft flybys to disk-integrated spectra and Earthshine, have demonstrated that atmospheric composition and surface-atmosphere interactions can encode information about habitability and biological activity \citep{Sagan1993,segura2005biosignatures,Kaltenegger2007,Robinson2011,Rugheimer2013,Rugheimer2015}. However, the modern atmosphere represents only the most recent state in a long evolutionary sequence. Over the past $\sim4$\,Gyr, Earth evolved from a largely anoxic world through prolonged partial oxygenation to the O$_2$-rich atmosphere observed today \citep{Holland2006,Lyons2014,CatlingZahnle2020,Olson2018}. Evidence for life on Earth extends back to at least $\sim3.8$\,Ga based on isotopic and geological signatures from some of the oldest terrestrial rocks \citep{mojzsis1996evidence,dodd2017evidence}, implying that inhabited conditions existed long before modern oxygen levels were established. In the future, temperate rocky exoplanets in habitable zones will be observed at unknown stages of geological and biological evolution, with some of these exoplanets following evolutionary trajectories that only partly resemble Earth's. Thus, interpreting every potentially habitable planet through a modern-Earth lens could cause us to miss inhabited worlds with little or no oxygen. It could also lead to false positives if abiotic gases are interpreted as signs of life without considering the broader planetary context \citep{Meadows2018,Schwieterman2018,Thompson2022}.

The simultaneous presence of an oxidized species such as O$_2$ or O$_3$ and a reduced species such as CH$_4$ is a long-standing example of atmospheric chemical disequilibrium as a possible biosignature \citep{Lederberg1965,Lovelock1965,Seager2012,Meadows2018,Schwieterman2018}. Its interpretation, however, depends strongly on atmospheric redox state and geochemical boundary conditions \citep{KrissansenTotton2018,KrissansenTotton2020}. Abiotic processes can generate O$_2$ and O$_3$ under some stellar and atmospheric conditions, CH$_4$ can have both biological and geological sources, and inhabited planets need not sustain remotely detectable modern-Earth O$_2$ abundances \citep{Meadows2018,Thompson2022}. Individual molecules are therefore insufficient in isolation; O$_2$ and CH$_4$ are most informative when considered together and with environmental tracers such as CO$_2$ and H$_2$O.

Earth's own atmospheric history illustrates this problem directly. The Archean atmosphere was predominantly anoxic, with CO$_2$ and CH$_4$ being the important greenhouse gases and with organic haze expected under sufficiently methane-rich conditions \citep{Trainer2006,Arney2016,CatlingZahnle2020}. Atmospheric O$_2$ began to rise during the Great Oxidation Event near $\sim2.4$--2.3\,Ga, but much of the Proterozoic remained weakly oxygenated with atmospheric redox chemistry influenced by CH$_4$, volcanic gases, and the sulfur cycle \citep{Holland2006,Lyons2014,Zahnle2006,CatlingZahnle2020}. A later rise in oxygen toward the Neoproterozoic preceded the modern high-O$_2$ atmosphere \citep{Lyons2014,Olson2018}. These stages span methane-rich anoxic atmospheres, intermediate states containing both reduced and oxygen-bearing species, and the strongly oxidized modern atmosphere. They provide useful evolutionary analogues for inhabited terrestrial planets without implying that all such planets must follow Earth's specific trajectory.

The observable consequences of these states also depend on the host star. Stellar ultraviolet radiation controls photolysis and therefore the steady-state abundances of species such as O$_3$ and CH$_4$ \citep{segura2005biosignatures,Rugheimer2015}, while stellar evolution changes the incident radiation field with time \citep{Claire2012}. Consequently, similar atmospheric boundary conditions can produce different chemical structures around different stellar hosts. Earth-through-time calculations have explored this dependence from both geological and stellar perspectives. \citet{Kaltenegger2007} modeled the spectral evolution of Earth across geological epochs, while subsequent studies extended these calculations across FGKM stars \citep{Rugheimer2013,Rugheimer2015}. In particular, \citet{Rugheimer2018} considered 3.9, 2.0, and 0.8\,Ga atmospheres together with modern Earth around an F0V--M8V grid, showing that molecular features depend jointly on geological state and host spectral type. Later high-resolution studies likewise demonstrated that combinations such as O$_2$+CH$_4$ and O$_3$+CH$_4$ can remain informative before modern oxygen abundances are reached \citep{Kaltenegger2020,Kaltenegger2020b,Kaltenegger2023}.

High-resolution spectroscopy provides a route for testing such atmospheric states observationally. At resolving powers of $R\sim10^5$, molecular bands are resolved into dense forests of spectral lines whose combined Doppler-shifted signal (at a few km\,s$^{-1}$) can be recovered from quasi-stationary stellar and telluric features through high-resolution cross-correlation spectroscopy (HRCCS) \citep{Snellen2013,Snellen2015,wyttenbach2015spectrally,brogi2016rotation,brogi2018exoplanet,Birkby2018,BrogiLine2019,sanchez2019water,giacobbe2021five,dubey2024comparative,dash2024constraints,dubey2025quantified,dash2026probing}. Temperate planets are particularly challenging because their radial velocities change only slowly during transit, increasing the susceptibility of the planetary signal to PCA- or SVD-based detrending. Injection--recovery studies nevertheless show that low-velocity signals can be recovered when sufficient out-of-transit spectra are available and the effect of detrending is propagated to the planetary template \citep{cheverall2024feasibility,dash2026probing}. However, stellar line-profile variations, including the Rossiter--McLaughlin effect \citep{mclaughlin1924some,rossiter1924detection}, provide an additional source of contamination because they can also produce phase-dependent residuals that can mimic a slow-moving exoplanetary signal, but whether the most common detrending processes used today for high-resolution spectroscopic analyses can remove this source of contamination still remains poorly understood. Nevertheless, the 39\,m Extremely Large Telescope (ELT) and its ArmazoNes high Dispersion Echelle Spectrograph (ANDES) will provide approximately $R\sim100{,}000$ spectroscopy from optical to near-infrared wavelengths, making temperate rocky planets a major target for high-resolution atmospheric characterization \citep{Palle2025}.

Previous studies illustrate the scale of the observational challenge. \citet{Snellen2013} obtained a 3.8$\sigma$ O$_2$ signal after 30 transits for an Earth analogue around an M5V star, while \citet{RodlerLopezMorales2014} estimated approximately 74 transits ($\sim175$\,h) for an M-dwarf system at 10\,pc. Using real ground-based spectra, \citet{SerindagSnellen2019} estimated about 20-50 transits for an optimal Proxima-Centauri-like case. Molecular detectability can also differ substantially: \citet{Currie2023} found requirements of approximately 34, 33, and 200 transits for CO$_2$, CH$_4$, and O$_2$ detections, respectively, in an Earth-like TRAPPIST-1\,e atmosphere. At the population level, \citet{HardegreeUllman2023} found that Earth-like O$_2$ could be accessible for only up to about 21\% of suitable nearby M-dwarf systems over a 50 year survey under their most favorable assumptions. More recently, \citet{KurzawaFerrandez2026} found H$_2$O to be the most accessible species, requiring about 10-19 transits for the TRAPPIST-1 planets and about 30 transits for LHS 1140\,b, whereas CO$_2$, CH$_4$, and O$_2$ required approximately 1.5, 3, and 4 times as many transits as H$_2$O, respectively. These studies establish the difficulty of high-resolution biosignature searches, but largely consider modern-Earth-like atmospheric compositions.

Here we instead ask how high-resolution molecular detectability changes as an Earth-like atmosphere evolves and as the host star changes. We consider two complementary model grids. The first explores solar-host Archean organic-haze chemistry, Proterozoic O$_2$--CH$_4$ states, and modern Earth. The second follows four geological epochs across an FGKM host-star grid motivated by \citet{Rugheimer2018}. We compute steady-state atmospheric profiles with \texttt{Photochem} \citep{Wogan2025Photochem}, generate high-resolution transmission spectra with \texttt{petitRADTRANS} \citep{Molliere2019,Molliere2020}, and evaluate the detectability of O$_2$, CH$_4$, H$_2$O, and CO$_2$ with the ANDES YJH bandpass using the \texttt{Ratri} simulation and cross-correlation framework \citep{dash2025detectability,dash2026probing}. Our aim is not to assign a unique observable spectrum to a geological age, but to determine how the molecular information available to high-resolution spectroscopy changes with atmospheric redox state and stellar environment, which combinations of species provide the strongest leverage between broad evolutionary states, and where optimistic photon-noise-limited ANDES predictions indicate the greatest observational potential. Section \ref{sec:methods} describes the atmospheric and observing framework, Section \ref{sec:results} presents the chemical, spectral, and detectability results, Section \ref{sec:discussion} discusses their interpretation and limitations, and Section \ref{sec:conclusion} summarizes the implications for future searches for inhabited terrestrial planets that may not resemble modern Earth.


\section{Methods}
\label{sec:methods}

\subsection{Atmospheric chemistry models}
\label{sec:chemistry}

We use the open-source \texttt{Photochem} package and its \texttt{EvoAtmosphere} solver to simulate the atmospheric chemistry. The code descends from the Atmos/Zahnle photochemical framework and solves coupled gas-phase chemistry, photolysis, transport, condensation, and particle microphysics on a one-dimensional vertical grid \citep{Zahnle2006,Kopparapu2013,Arney2016}. Atmospheric chemistry spans over 100 atmospheric layers in all our models. For each case, the temperature profile and eddy-diffusion profile are prescribed and kept fixed while the chemistry is integrated to steady state. A fully coupled climate–chemistry iteration is beyond the scope of the present study and is deferred to future work.

\begin{table}
\caption{Archean H1 photochemical grid following \citet{Arney2016}. All cases use a constant $p\mathrm{O}_2=10^{-8}$\,bar.}
\label{tab:archean_grid}
\centering
\begin{tabular}{lccccc}
\hline
Case & $P_{\rm surf}$ & $p\mathrm{CO}_2$ & CH$_4$/CO$_2$ & $p\mathrm{CH}_4$ & $T_{\rm surf}$ \\
 & (bar) & (bar) &  & (bar) & (K) \\
\hline
A1-H1 & 1.0 & 0.0100 & 0.15 & 0.00150 & 284 \\
A2-H1 & 1.0 & 0.0100 & 0.19 & 0.00190 & 263 \\
A3-H1 & 1.0 & 0.0100 & 0.26 & 0.00260 & 263 \\
B1-H1 & 1.0 & 0.0180 & 0.10 & 0.00180 & 299 \\
B2-H1 & 1.0 & 0.0180 & 0.17 & 0.00306 & 274 \\
B3-H1 & 1.0 & 0.0180 & 0.21 & 0.00378 & 274 \\
C1-H1 & 0.5 & 0.0100 & 0.15 & 0.00150 & 282 \\
C2-H1 & 0.5 & 0.0100 & 0.18 & 0.00180 & 282 \\
C3-H1 & 0.5 & 0.0100 & 0.21 & 0.00210 & 262 \\
D1-H1 & 0.5 & 0.0036 & 0.20 & 0.00072 & 273 \\
D2-H1 & 0.5 & 0.0036 & 0.28 & 0.001008 & 251 \\
D3-H1 & 0.5 & 0.0036 & 0.31 & 0.001116 & 251 \\
\hline
\end{tabular}
\end{table}

\subsubsection{Archean grid}
\label{sec:methods:archean}

The Archean models follow the organic-haze (H1) climate families of \citet{Arney2016}. The grid spans four background climate states (A--D) and three CH$_4$/CO$_2$ ratios within each state, yielding 12 atmospheres (see Table \ref{tab:archean_grid}). CO$_2$ and CH$_4$ are imposed through fixed lower-boundary partial pressures, O$_2$ is fixed at $10^{-8}$\,bar, and N$_2$ fills the remaining dry atmosphere. We use the chemical network \texttt{zahnle\_HNOCS.yaml}, which includes hydrocarbon chemistry and organic-haze precursors, for our study.  Following Supplementary Table 2 of \citet{Arney2016}, lower boundary conditions combine surface deposition with volcanic outgassing, where both apply. H$_2$, SO$_2$, and H$_2$S are supplied at volcanic fluxes of $3.5\times10^{9}$, $10^{10}$, and $3.5\times10^{8}$\,molecules\,cm$^{-2}$\,s$^{-1}$ together with deposition velocities of $2.4\times10^{-4}$, $1.0$, and $0.021$\,cm\,s$^{-1}$, respectively. CO is removed by deposition alone at $1.2\times10^{-4}$\,cm\,s$^{-1}$. Additional deposition velocities are imposed for O, H, OH, HO$_2$, HCO, CH$_3$, HNO, HSO, H$_2$SO$_4$, H$_2$O$_2$, H$_2$CO, HNO$_3$, NO, NO$_2$, O$_3$ and the condensed particles S$_8$, H$_2$SO$_4$, and HCaer1 (for further details, see Supplementary Table S2 from \citet{Arney2016}). Water vapor is maintained with an ocean reservoir and is allowed to condense onto H$_2$O$\mathrm{_{aer}}$ at a relative humidity of $0.8$. Case-to-case changes are restricted to the fixed surface pressures of N$_2$, CO$_2$, and CH$_4$ (with O$_2$ held at $10^{-8}$\,bar in all Archean H1 runs). The initial pressure--temperature profiles are generated with \texttt{AdiabatClimate} \citep{Wogan2025Photochem}, using the haze-bearing surface temperatures reported by \citet{Arney2016}, an isothermal stratosphere at $180$\,K, and $K_{zz}=10^{6}$\,cm$^{2}$\,s$^{-1}$.

\begin{table}
\caption{Proterozoic V2 photochemical grid from \citet{Zahnle2006}. The background atmosphere has $p\mathrm{CO}_2=0.01$\,bar and $T_{\rm surf}=284$\,K.}
\label{tab:proterozoic_grid}
\centering
\begin{tabular}{lccc}
\hline
Cases & $f(\mathrm{O}_2)$ & $f(\mathrm{CH}_4)$ & Notes \\
\hline
A1-V2 & $10^{-10}$ & $10^{-6}$ & very anoxic, low CH$_4$ \\
A2-V2 & $10^{-10}$ & $10^{-5}$ &  \\
A3-V2 & $10^{-10}$ & $10^{-4}$ &  \\
A4-V2 & $10^{-10}$ & $10^{-3}$ &  very anoxic, CH$_4$-rich\\
B1-V2 & $10^{-7}$ & $10^{-6}$ & weakly oxygenated, low CH$_4$ \\
B2-V2 & $10^{-7}$ & $10^{-5}$ &  \\
B3-V2 & $10^{-7}$ & $10^{-4}$ &  \\
B4-V2 & $10^{-7}$ & $10^{-3}$ &  weakly oxygenated, CH$_4$-rich\\
C1-V2 & $10^{-3}$ & $10^{-6}$ & oxic, low CH$_4$ \\
C2-V2 & $10^{-3}$ & $10^{-5}$ &  \\
C3-V2 & $10^{-3}$ & $10^{-4}$ &  \\
C4-V2 & $10^{-3}$ & $10^{-3}$ & oxic, CH$_4$-rich \\
\hline
\end{tabular}
\end{table}

\begin{figure}[h!]
\centering
\includegraphics[width=\columnwidth]{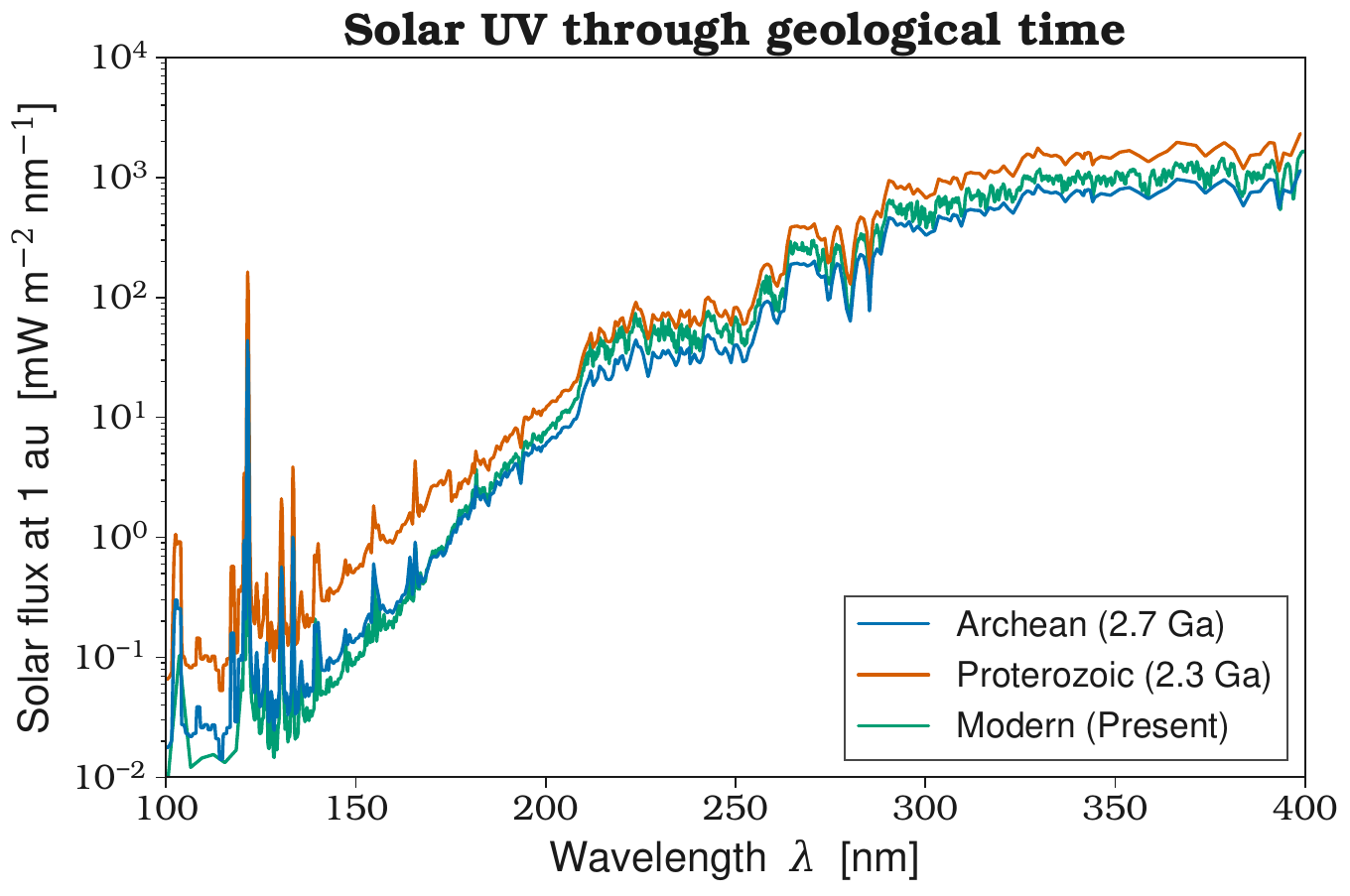}
\caption{Solar ultraviolet spectra implemented for the Archean, Proterozoic, and modern solar-host calculations. The curves follow the separate stellar prescriptions used for the three model families described in Sections \ref{sec:methods:archean} and \ref{sec:methods:proterozoic}.}
\label{fig:solar_uv_eras}
\end{figure}

The incident spectrum is a 2.7\,Ga solar spectrum based on the wavelength-dependent solar evolution prescription of \citet{Claire2012}. This age follows the late-Archean framework of \citet{Arney2016}, who adopted a 2.7\,Ga Sun to match the age of the geochemical CO$_2$ constraints used to define their atmospheric scenarios \citep{driese2011neoarchean}. It should therefore be regarded as representative of a late-Archean atmosphere. Because solar evolution is wavelength dependent, an earlier Archean Sun would differ most strongly at short UV wavelengths relevant to photolysis and haze-producing chemistry; exploring this temporal dependence across the full Archean is beyond the scope of the present grid. We use a surface albedo of 0.32, a solar zenith angle of $60^\circ$, and diffusion-limited hydrogen escape. The resulting grid is intended to follow the transition from comparatively haze-poor to strongly methane-driven hydrocarbon chemistry while retaining the background climate states explored by \citet{Arney2016}.

\subsubsection{Proterozoic grid}
\label{sec:methods:proterozoic}

The Proterozoic models are constructed around the two principal redox controls explored by \citet{Zahnle2006}: surface O$_2$ and CH$_4$. We use a 1\,bar N$_2$ background with $p\mathrm{CO}_2=0.01$\,bar and the ``modern high'' (V2) volcanic sulfur fluxes. Surface O$_2$ mixing ratios are $10^{-10}$, $10^{-7}$, and $10^{-3}$, while CH$_4$ mixing ratios are $10^{-6}$, $10^{-5}$, $10^{-4}$, and $10^{-3}$. The resulting 12 models are listed in Table \ref{tab:proterozoic_grid}. Organic haze is disabled so that changes in the gas-phase oxidation state can be isolated.

The thermal structure of the atmosphere follows the representative setup of \citet{Zahnle2006}, with $T_{\rm surf}=284$\,K, a tropopause near 11\,km, and an isothermal 186\,K stratosphere. The eddy diffusion coefficient is $10^{5}$\,cm$^2$\,s$^{-1}$ in the troposphere and decreases with pressure in the stratosphere following the Zahnle prescription. We use a solar zenith angle of $50^\circ$, a surface albedo of 0.32, and relative humidity of 0.8. The stellar spectrum represents a 2.3\,Ga Sun with the ultraviolet enhancement adopted by \citet{Zahnle2006}: a factor of four shortwards of 175\,nm and two from 175--400\,nm relative to the modern spectrum, together with the solar-age scaling of \citet{Claire2012} and \citet{Ribas2005}.

\begin{figure*}
\centering
\includegraphics[width=0.78\textwidth]{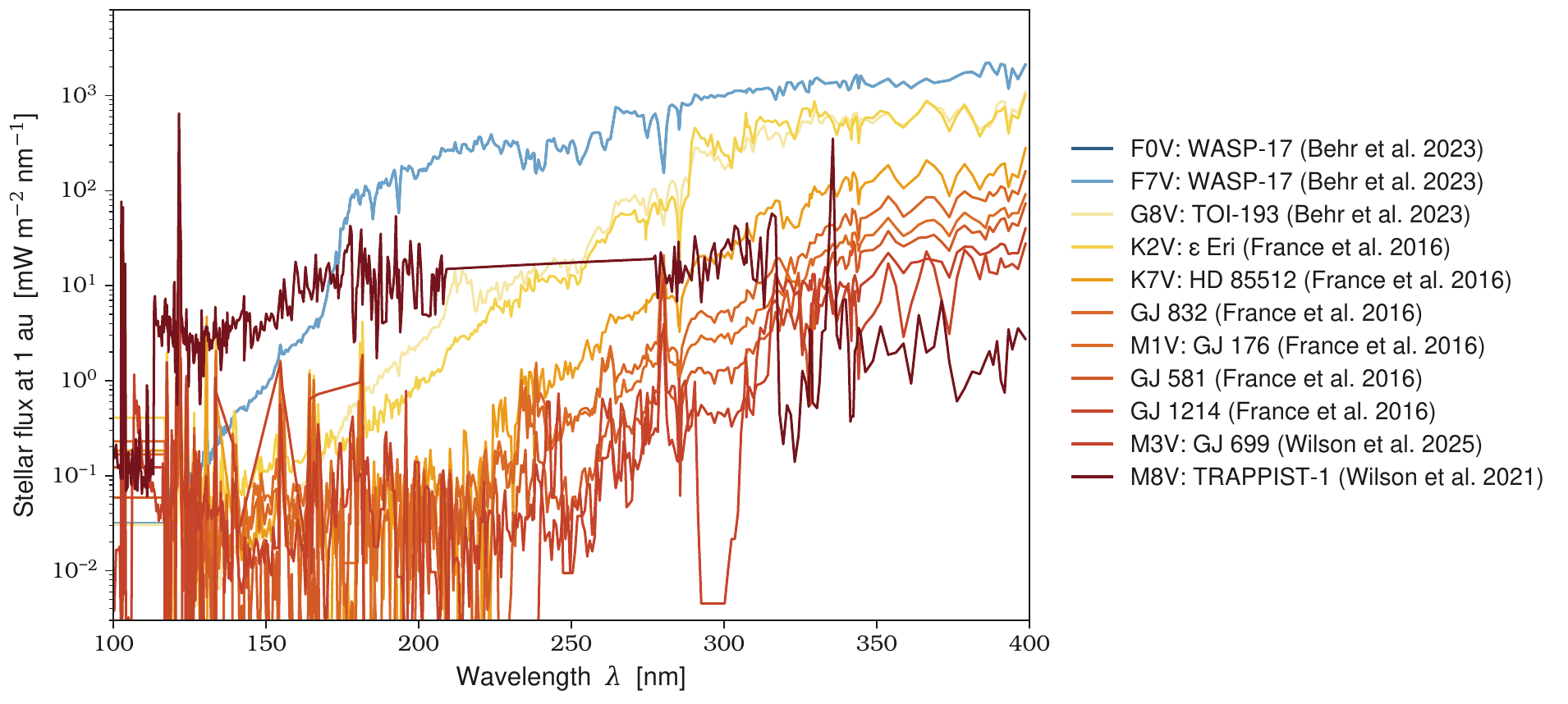}
\caption{Ultraviolet spectral energy distributions for the multi-star grid. Each stellar proxy is renormalized to the epoch-dependent bolometric irradiation used by the photochemical model. The wavelength-dependent UV shape therefore varies with host star and directly controls photolysis.}
\label{fig:grid_uv_seds}
\end{figure*}

\subsubsection{Modern Earth benchmark}
\label{sec:methods:modern}

The modern benchmark uses the \texttt{zahnle\_earth.yaml} network and the present-day solar spectrum distributed with \texttt{Photochem} (the comparison of the solar spectrum across three ages are shown in Figure \ref{fig:solar_uv_eras}). The lower boundary fixes $p\mathrm{N}_2=0.78$\,bar, $p\mathrm{O}_2=0.21$\,bar, $p\mathrm{CO}_2=3.6\times10^{-4}$\,bar, and trace abundances of H$_2$, CO, and N$_2$O, while CH$_4$ is supplied at $9\times10^{10}$ molecules\,cm$^{-2}$\,s$^{-1}$. SO$_2$ and H$_2$S are included at fluxes of $10^{9}$ and $10^{8}$ molecules\,cm$^{-2}$\,s$^{-1}$, respectively. Rainout and species-dependent dry deposition are included for soluble and reactive species. We keep a solar zenith angle of $60^\circ$, a surface albedo of 0.25, a near-surface $K_{zz}=10^{5}$\,cm$^2$\,s$^{-1}$, and water condensation at a relative humidity of 0.5. The converged surface temperature is approximately 286.5\,K.

\subsubsection{FGKM Earth-through-time grid}
\label{sec:methods:rugheimer}

To examine the effect of the stellar environment, we construct a second atmospheric grid following the design of \citet{Rugheimer2018}. Four epochs are considered across stellar hosts: 3.9\,Ga, 2.0\,Ga, 0.8\,Ga, and modern Earth. The host grid contains F0V, F7V, G8V, K2V, K7V, M1V, M3V, and M8V spectral types, together with GJ\,581, GJ\,832, and GJ\,1214, yielding 44 atmosphere-host combinations. The proxy star does not necessarily belong to the same nominal spectral subtype for all the host star cases, because suitable UV spectra are not available for every spectral type in the MUSCLES databases. Where an exact match is unavailable, we use the closest suitable stellar proxy with available UV coverage.

At a given epoch, each host delivers the same bolometric irradiation to the model atmosphere, as in \citet{Rugheimer2018}. The top-of-atmosphere flux is
\begin{equation}
F_{\star}=1370 f_L\quad \mathrm{W\,m^{-2}},
\end{equation}
where $f_L=0.746$, 0.852, 0.935, and 1.0 for 3.9\,Ga, 2.0\,Ga, 0.8\,Ga, and modern Earth, respectively \citep{Claire2012}. The wavelength dependence of the ultraviolet radiation is taken from MUSCLES, Mega-MUSCLES, or MUSCLES Extension proxy spectra \citep{France2016,Wilson2021,Behr2023,Wilson2025}. The proxy spectra are renormalized to the appropriate epoch-integrated flux before being supplied to \texttt{Photochem}. The final host-star UV spectra are shown in Figure \ref{fig:grid_uv_seds}. 

Surface CO$_2$ and O$_2$ are fixed to the epoch-dependent values of \citet{Rugheimer2018}, while the CH$_4$ and N$_2$O lower-boundary abundances follow their host-dependent values. Surface temperatures \(T_{\mathrm{surf}}\) for the F0V-M8V grid points are taken from the top panel of Figure 2 of \citet{Rugheimer2018}. For the additional MUSCLES hosts GJ\,581, GJ\,832, and GJ\,1214, we assign \(T_{\mathrm{surf}}\) at each epoch by linear interpolation in effective temperature between the M3V host (\(T_{\mathrm{eff}}=3200\,\mathrm{K}\)) and M1V host (\(T_{\mathrm{eff}}=3700\,\mathrm{K}\)) anchors,
\begin{equation}
T_{\mathrm{surf}}(T_{\mathrm{eff}})
=
T_{\mathrm{surf}}^{\mathrm{M3V}}
+
\frac{T_{\mathrm{eff}}-3200\,\mathrm{K}}{500\,\mathrm{K}}
\bigl(
T_{\mathrm{surf}}^{\mathrm{M1V}}
-
T_{\mathrm{surf}}^{\mathrm{M3V}}
\bigr),
\end{equation}
and round the result to the nearest kelvin. The host effective temperatures used for this interpolation are \(T_{\mathrm{eff}}=3498\,\mathrm{K}\) (GJ\,581), \(3620\,\mathrm{K}\) (GJ\,832), and \(3250\,\mathrm{K}\) (GJ\,1214). The models use a 1\,bar dry atmosphere, Earth mass and radius, a surface albedo of 0.32, a solar zenith angle of $60^\circ$, and $K_{zz}=10^{5}$\,cm$^2$\,s$^{-1}$. The 3.9\,Ga cases use the H-N-O-C-S network, whereas the oxygenated epochs use the modern-Earth network. We do not impose organic haze in this grid.

\begin{table*}
\caption{Stellar hosts used for the detectability calculations. Period ($P$) and transit duration ($T_{14}$) correspond to an Earth twin on a circular Earth-insolation orbit with impact parameter $b=0.3$. Exposure times are those used in the \texttt{Ratri} simulations for each stellar host.}
\label{tab:host_observing}
\centering
\begin{tabular}{llcccccl}
\hline
Host & Proxy & $R_\star$ & $d$ & $P$ & $T_{14}$ & $t_{\rm exp}$ & SED source \\
 &  & ($R_\odot$) & (pc) & (days) & (hours) & (s) &  \\
\hline
F0V & WASP-17 & 1.57 & 403 & 898.2 & 23.81 & 69.0 & \citet{Behr2023} \\
F7V & WASP-17 & 1.57 & 403 & 898.2 & 23.81 & 69.0 & \citet{Behr2023} \\
Sun & -- & 1.00 & 150 & 365.25 & 12.50 & 30.0 & Solar$^c$ \\
G8V & TOI-193 & 0.95 & 80.44 & 293.3 & 11.14 & 10.5 & \citet{Behr2023} \\
K2V & $\varepsilon$ Eri & 0.76 & 3.212 & 175.7 & 7.97 & 0.5$^a$ & \citet{France2016} \\
K7V & HD\,85512 & 0.71$^b$ & 11.28 & 124.2 & 7.17 & 0.5 & \citet{France2016} \\
M1V & GJ\,176 & 0.48 & 9.485 & 45.0 & 3.79 & 1.4 &  \citet{France2016} \\
M3V & GJ\,699 & 0.18 & 1.827 & 15.4 & 1.80 & 0.55 &  \citet{Wilson2025} \\
M8V & TRAPPIST-1 & 0.12 & 12.47 & 4.41 & 0.82 & 90.5 &  \citet{Wilson2021} \\
GJ\,581 & GJ\,581 & 0.32 & 6.300 & 25.6 & 2.45 & 1.6 &  \citet{France2016} \\
GJ\,832 & GJ\,832 & 0.48 & 4.967 & 47.5 & 4.10 & 0.5$^a$ &  \citet{France2016} \\
GJ\,1214 & GJ\,1214 & 0.22 & 14.64 & 13.4 & 1.65 & 25.5 &  \citet{France2016} \\
\hline
\end{tabular}
\tablecomments{$^a$The exposure-time floor places these nearby systems near the soft-saturation regime.}
\tablecomments{$^b$The MUSCLES radius is unavailable; we took $R_\star=0.71R_\odot$ from \citet{Stassun2017}.}
\tablecomments{$^c$The solar-host atmospheric calculations use the epoch-dependent solar spectra described in Sections \ref{sec:methods:archean}--\ref{sec:methods:modern}.}
\end{table*}

\subsection{High-resolution transmission spectra}
\label{sec:spectra}

For each converged atmospheric column, we compute line-by-line radiative transfer and obtain clear-atmosphere transmission spectra with \texttt{petitRADTRANS}. The pressure, temperature, mean molecular weight, and volume mixing-ratio profiles from \texttt{Photochem} are mapped onto the radiative-transfer grid. All planets are assigned $R_{\rm p}= 1\,R_\oplus$ and $M_{\rm p}=1\,M_\oplus$, with a reference radius specified at 0.01\,bar. Spectra are calculated from 0.4--1.8\,$\mu$m at an effective resolving power of $10^5$.

The opacity budget includes CO$_2$, CH$_4$, CO, HCN, C$_2$H$_6$, H$_2$O, O$_2$, and O$_3$ where present; the multi-star grid also includes N$_2$O. In all simulations we include Rayleigh scattering by N$_2$, O$_2$, H$_2$O, CO$_2$, and H$_2$ whenever those species are present in the Photochem abundance profiles. N$_2$ is the dominant background gas in these Earth-like atmospheres, so N$_2$ Rayleigh sets most of the continuum baseline; the remaining Rayleigh partners are retained because O$_2$, H$_2$O, and CO$_2$ can contribute non-negligibly when their abundances rise (oxygenated eras; moist or CO$_2$-rich cases), while H$_2$ Rayleigh is included for consistency in H$_2$-bearing Archean mixtures. The adopted line data are based on HITEMP and HITRAN databases: H$_2$O, CO$_2$, and CO from HITEMP \citep{HITEMP2010}; CH$_4$ from \citet{Hargreaves2020}; HCN from \citet{harris2002opacity}; C$_2$H$_6$ uses HITRAN2024 \citep{gordon2026hitran2024}, and N$_2$ from the WCCRMT database \citep{western2018spectrum}. The high-resolution O$_2$ \citep{gordon2022hitran2020}, O$_3$ \citep{rothman2013hitran2012}, and N$_2$O \citep{hargreaves2019spectroscopic} tables were obtained from DACE opacity database \citep{grimm2021helios} and converted into the \texttt{petitRADTRANS} format. The O$_2$ opacity includes the 0.76\,$\mu$m A-band, whereas our O$_3$ table begins at 1.4\,$\mu$m. 

Organic haze and clouds are intentionally omitted from the radiative transfer. We also neglect atmospheric refraction for our temperate Earth-twin transit geometry; under these conditions gas and Rayleigh opacity dominate the slant optical depth, so refraction remains as a minor contributor to the continuum \citep{robinson2017analytic,betremieux2013transmission}. The resulting spectra should therefore be interpreted as gas-only limiting cases for HRCCS. Haze is retained in the Archean photochemistry because it affects the chemical state of the atmosphere, but its continuum extinction is not propagated into the transmission spectrum.


\subsection{ELT/ANDES simulations with \texttt{Ratri}}
\label{sec:ratri}

The generated high-resolution transmission spectra are used as inputs to the open-source \texttt{Ratri}\footnote{https://github.com/dashspandan/Ratri} high-resolution observational simulation framework to estimate their detectabilities using ELT/ANDES. \texttt{Ratri} itself has already been validated against the current version of the ANDES ETC simulator across all bands in \citet{dash2026probing} for its current configuration, but in this study we retain only the YJH band with CARMENES CCD configuration used in \citet{dash2026probing} and generate a noisy exposure (with a corresponding signal-to-noise per pixel or $S/N_{n_\mathrm{spectra}=1}$, see Figure \ref{fig:snr_host}) incorporating wavelength-dependent stellar photon flux, telluric absorption through the Earth's atmosphere, instrumental response tailored to CARMENES CCDs from \citet{gandhi2020seeing}, and a Poissonian noise treatment for calculating the  signal-to-noise (S/N) per pixel. The errors at all pixels are assumed to be uncorrelated, and thus the actual noise at each pixel is randomly drawn from a normal distribution with variance equivalent to the photon flux error value calculated at that pixel assuming Poisson statistics. Since we anticipate the difficulty of detecting our generated spectra for almost all stellar hosts within a few transits, we use the time-invariant form of \texttt{Ratri} using the \texttt{ratri\_1D.txt} script. To simulate telluric absorption, we use an airmass value of 1.2 and a precipitable water vapor (PWV) value between 2.5-3.5\,mm as inputs for the ESO SkyCalc module \citep{noll2012atmospheric, jones2013advanced} integrated within \texttt{Ratri}. 

The time-invariant script injects the exoplanetary signal at one particular value of exoplanetary radial velocity (RV) for an exposure (away from 0\,kms$^{-1}$ where any residual telluric contamination can artificially enhance the signal due to artifacting), and then stacks $n_\mathrm{spectra}$ exposures together to enhance the total $S/N$ ratio per pixel through the relation:

\begin{equation}
    \bigg{(}\frac{S}{N}\bigg{)}_{n_\mathrm{spectra}} = \sqrt{n_\mathrm{spectra}}\bigg{(}\frac{S}{N}\bigg{)}_{n_\mathrm{spectra}=1} = \frac{S_{n_\mathrm{spectra}=1}}{\frac{N_{n_\mathrm{spectra}=1}}{\sqrt{n_\mathrm{spectra}}}}. \label{dampsnr}
\end{equation}
The above relation follows from the fact that the total signal $\propto$ n$_\mathrm{spectra}$, with the total noise $\propto$ $\sqrt{n_\mathrm{spectra}}$ assuming Poisson statistics. The factor of $\sqrt{n_\mathrm{spectra}}$ used to damp down the noise for a single exposure in the last term of the equation is referred to as the \emph{damping factor}, which carries information about the total number of exposures stacked to get the corresponding $S/N$. 

After damping down the noise, we perform an idealized detrending step. The noisy exposure is divided by a simulation of the same exposure without noise or an injected planetary signal. This leaves a residual array containing only the noise and the injected planetary signal at a given RV. Here, we also mask the pixels S/N $\leq$ 50 as a low-flux correction step to remove the contribution of pixels that would realistically have the most telluric contamination and would also be masked out in real observations as well. Subsequently, a search for the injected signal is carried out by Doppler-shifting a model spectral template across a grid of RVs (on to the same pixel array as the residual dataset) and then cross-correlating with the residual array, with the same pixels also masked for the Doppler-shifted model array as in the residual array calculated before. If the peak of the cross-correlation function (CCF) stands out appreciably at some value of RV in comparison to its neighbouring values, we count it as a detection. To quantify the significance of detection of the peak of CCF compared to the rest of the RV-grid after cross-correlation, we use the CCF-to-logL (log-Likelihood) equation of \citet{BrogiLine2019}:

\begin{equation}
        \log(L) = -\frac{N}{2}\log[s_{f}^{2} - 2R(s) + s_{g}^{2}], \label{logleq}
\end{equation}

where $s_{f}^{2}$ (data variance), $s_{g}^{2}$ (model variance) and $R(s)$ (cross-covariance) are calculated as
\begin{eqnarray}
s_{f}^{2} & = & \frac{1}{N}\sum_{n}f^{2}(n), \nonumber \\
s_{g}^{2} & = & \frac{1}{N}\sum_{n}g^{2}(n), \nonumber \\
R(s)      & = & \frac{1}{N}\sum_{n}f(n)g(n-s). 
\end{eqnarray}
$f(n)$ and $g(n)$ are found by subtracting the mean from all unmasked values in each row of each spectroscopic order in the detrended residual array, and the corresponding row from the Doppler-shifted model array respectively. From here, we use a 1D version (with RV as the sole free-parameter) of the $\sigma$-calculation criterion based on Wilks' theorem \citep{wilks1938large} used in \citet{dash2024constraints}. We note that this $\sigma$ calculation is based on likelihoods and is different from the CCF-based signal-to-noise ratio (SNR) metric used in many other literature. For our metric, a signal confidently detected will have a value of 0-$\sigma$ and the rest of the grid will have values $>4-5\sigma$, thus meaning that our analysis is able to confidently reject most of the tested RV-grid at $>4-5\sigma$ compared to the injected RV solution. 

Our detection metric differs from the evidence-based approach of \citet{KurzawaFerrandez2026}. They compute Bayesian evidences, $\log (Z)$, using Nested Sampling and compare a signal model, in which the data are cross-correlated with Doppler-shifted planetary templates, against a null model (no injected planetary signal) through a Bayes factor test. Our approach thus tests how significant the signal at the RV value corresponding to the CCF peak is compared with the CCF values across the rest of the RV grid. Rather than testing whether the signal itself can be distinguished from the cross-correlation obtained from a dataset without any injected signal, it evaluates the significance of the recovered peak relative to the surrounding RV grid. However, we note that spurious residuals mimicking a signal rarely go above a 3$\sigma$ level with our approach and generally lie away from the injected RV values (see the bottom panel in Figure \ref{fig:ccf_m8v_ch4} for the 3.9 Ga case). Hence, we prefer to use a much tighter 5$\sigma$ threshold to count true detections to limit the possibility of a spurious signal leading to false detections of molecules for our grid of models.

For the solar-host detectability calculations, we use a Sun-like star with $T_{\rm eff}=5772$\,K, $R_\star=1R_\odot$, and $M_\star=1M_\odot$. The system is placed at a fiducial distance of 150\,pc, for which an exposure time of 30\,s keeps the continuum count level close to $7\times10^4$ electrons per resolution element while avoiding saturation of the CCD pixels. This operating point remains below the adopted soft-saturation limit and is conservative relative to the $>70000$\,e$^-$ per-pixel well capacity reported for the Teledyne HAWAII-4RG-15 (2.5\,\micron\ cutoff) detectors baselined for ANDES YJH \citep{zandian2016performance}. To estimate the number of spectra that can be obtained within a given observing interval, and hence the value of $n_{\rm spectra}$ used in Equation \ref{dampsnr}, we assume a 34\,s readout time between successive exposures, following the CARMENES timing currently implemented in \texttt{Ratri} \citep{Quirrenbach2014,dash2026probing}. This is slightly more conservative than the 30\,s overhead stated by \citet{Palle2025}. Using the same Earth-twin orbital prescription and impact parameter of $b=0.3$ described in Section \ref{sec:methods:orbit}, Equation \ref{eq:t14} gives a transit duration of $T_{14}\simeq12.5$\,h. This duration is used to convert the required number of simulated frames into transits and equivalent observing nights. 

For each stellar proxy in the FGKM grid, the exposure time is similarly adjusted to keep the continuum count level close to $7\times10^4$ electrons per resolution element while avoiding saturation of the CCD pixels, subject to a minimum exposure time of 0.5\,s.  For the nearest K- and M-dwarf proxies, the 0.5\,s exposure-time floor places some simulations close to the soft-saturation regime, and the corresponding detectability estimates should therefore be interpreted with caution.

To enable comparisons of detectability, we standardize the calculated sigma to that obtained from a stacking of $N_{\rm ref}=1600$ simulated frames. For each atmosphere--host-star combination, we measure the corresponding detection significance, $\sigma_{\rm ref}$ at $N_{\rm ref}=1600$. In the photon-noise-dominated regime, this significance is expected to scale approximately as $\sqrt{N}$ with the number of combined exposures \citep{Snellen2013,Birkby2018}. We therefore estimate the number of frames ($N_{\rm frames}$) required to reach a target significance ($\sigma_t$) as

\begin{equation}
N_{\rm frames} = N_{\rm ref}\left(\frac{\sigma_t}{\sigma_{\rm ref}}\right)^2.
\label{eq:nframes}
\end{equation}

Unless stated otherwise, we adopt $\sigma_t=5$ to correspond to a strong detection. We do not impose a minimum significance threshold on the 1600-frame reference calculation. For every finite and positive value of $\sigma_{\rm ref}$, the number of frames required to reach the target significance is estimated using the $\sqrt{N}$ scaling described above and converted into the corresponding number of transits and equivalent observing nights. Configurations requiring more than 500 transits or 200 equivalent observing nights are considered not practically detectable within the observing limits and are denoted as n.d. (non-detection). Cases with missing, non-finite, or non-positive values of $\sigma_{\rm ref}$ are treated similarly.
\begin{figure*}
\centering
\begin{minipage}[b]{0.49\textwidth}\includegraphics[width=\textwidth]{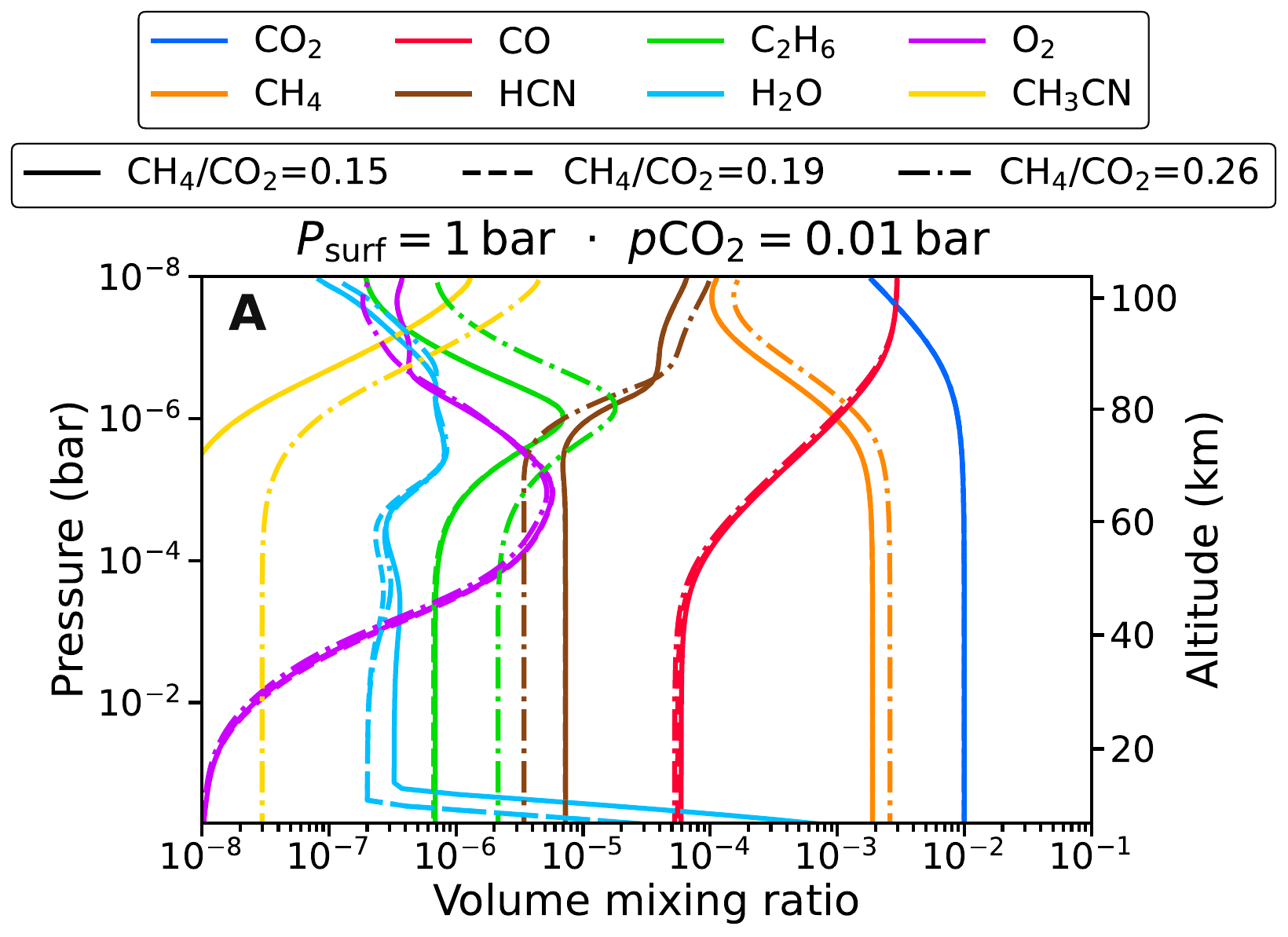}\end{minipage}
\begin{minipage}[b]{0.49\textwidth}\includegraphics[width=\textwidth]{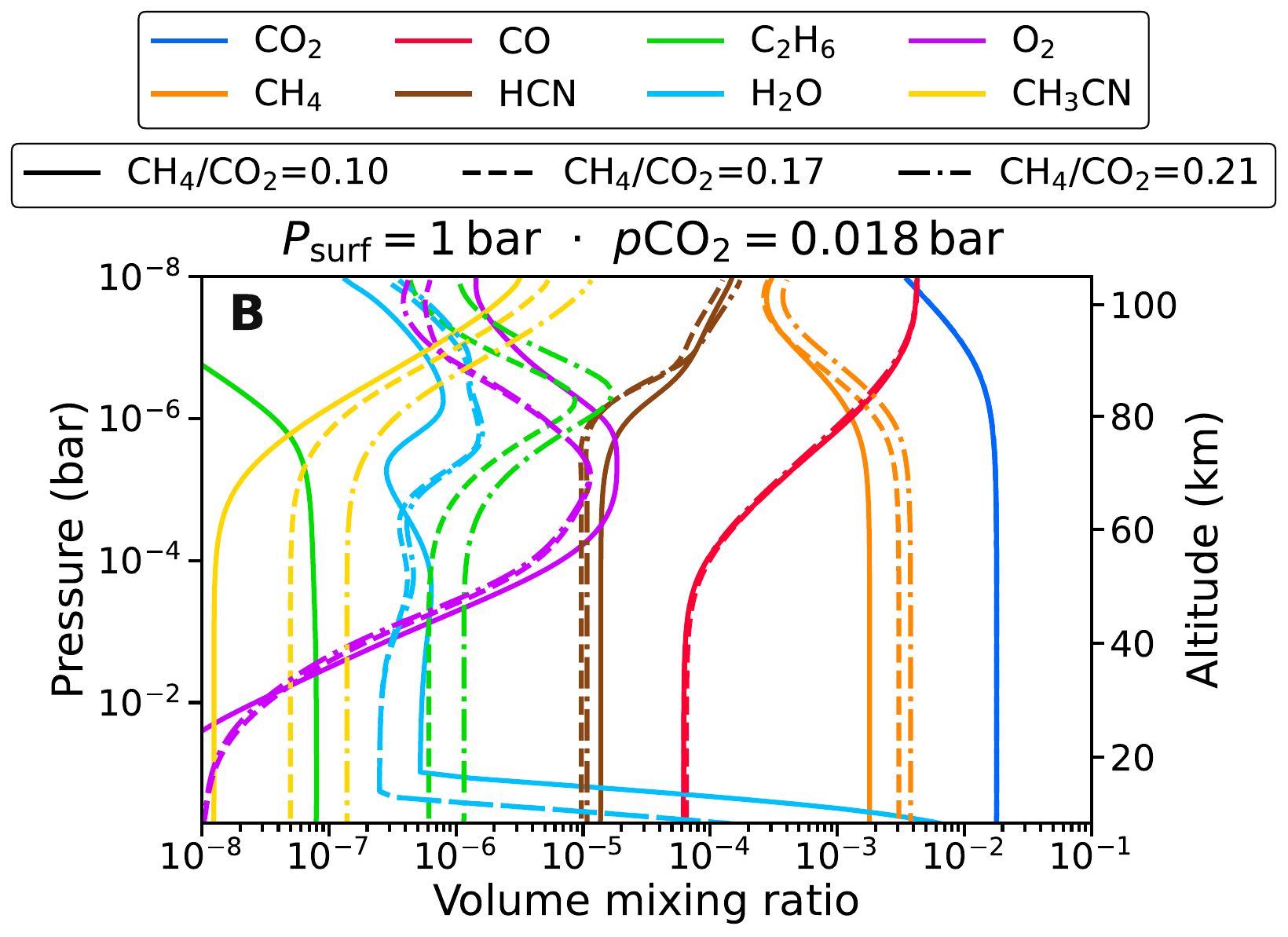}\end{minipage}\\[0.6cm]
\begin{minipage}[b]{0.49\textwidth}\includegraphics[width=\textwidth]{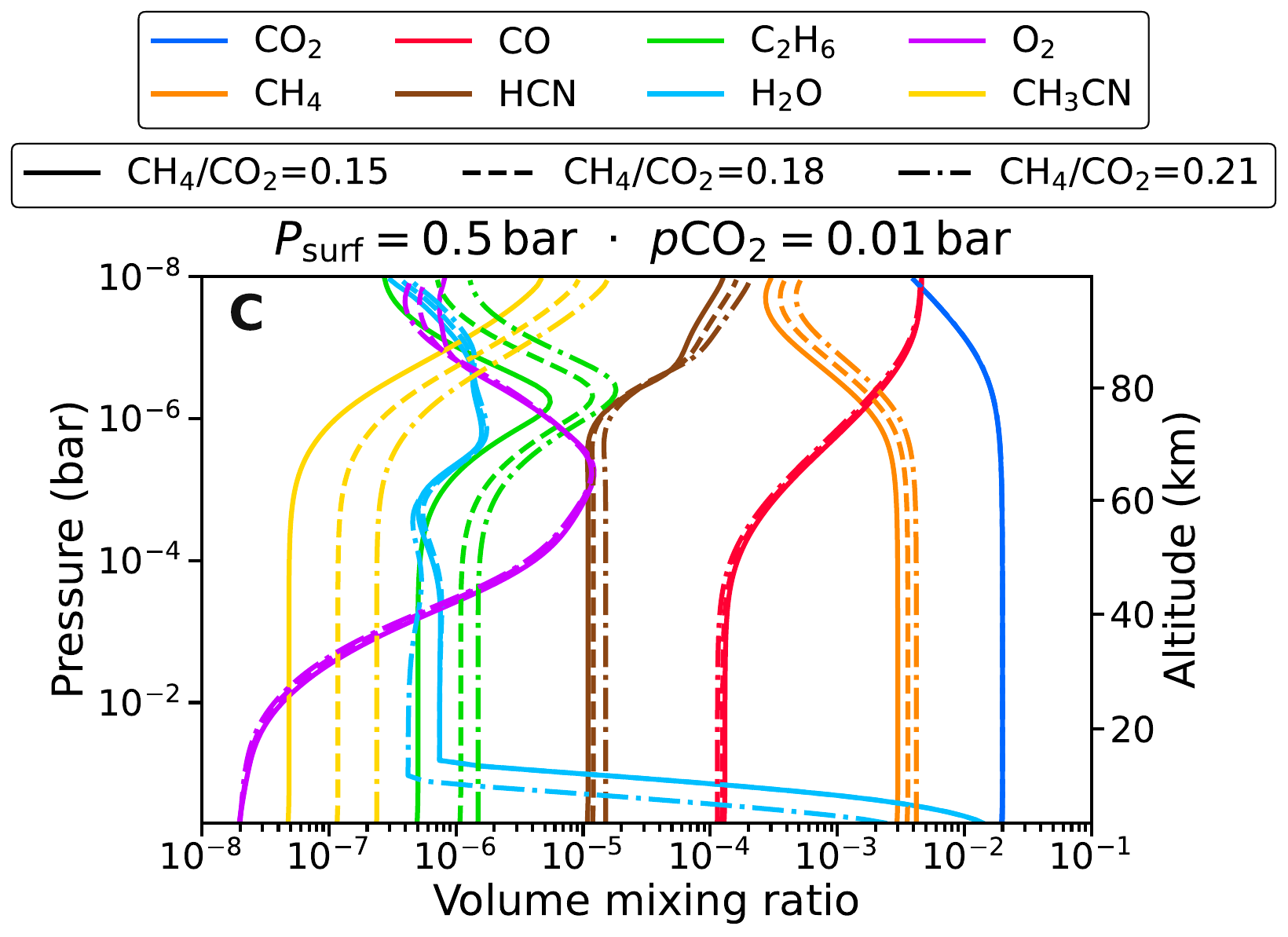}\end{minipage}
\begin{minipage}[b]{0.49\textwidth}\includegraphics[width=\textwidth]{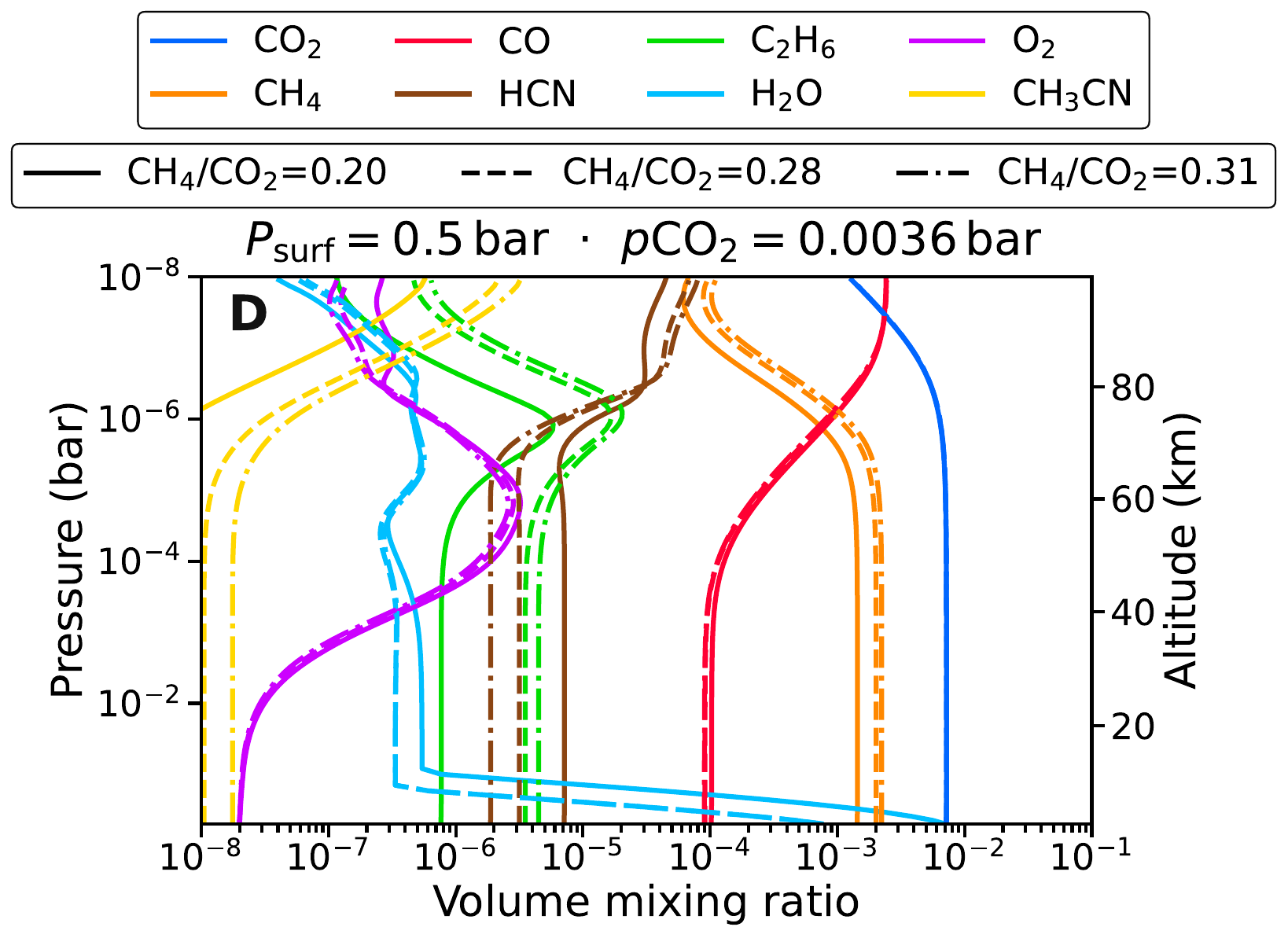}\end{minipage}
\caption{Steady-state mixing-ratio profiles for the Archean H1 grid. The four panels correspond to the A--D climate families from Table \ref{tab:archean_grid}; within each panel, increasing CH$_4$/CO$_2$ drives progressively stronger hydrocarbon chemistry.}
\label{fig:archean_chemisrty}
\end{figure*}

\subsubsection{Orbital and observing assumptions}
\label{sec:methods:orbit}

For the multi-star grid, each atmospheric model is assigned to an Earth twin with $R_{\rm p}=1\,R_\oplus$ and $M_{\rm p}=1\,M_\oplus$ on a circular orbit. The planet is placed at an orbital distance where the total incident bolometric flux equals that received by present-day Earth. This choice provides a common irradiation reference across the stellar grid and is not intended to define the planet as lying within the habitable zone, although the resulting orbit may overlap with conventional habitable-zone limits for some stellar types. The stellar luminosity ($L\mathrm{_{\star}}$) is calculated from the proxy radius ($R\mathrm{_{\star}}$) and effective temperature ($T\mathrm{_{eff}}$),

\begin{equation}
\frac{L_\star}{L_\odot}=\left(\frac{R_\star}{R_\odot}\right)^2\left(\frac{T_{\rm eff}}{T_{{\rm eff},\odot}}\right)^4,
\end{equation}

where $T_{{\rm eff},\odot}=5772$\,K. The stellar mass ($M\mathrm{_{\star}}$) is estimated from the radius and surface gravity (log $g$) used in the simulation,

\begin{equation}
\frac{M_\star}{M_\odot}=\left(\frac{R_\star}{R_\odot}\right)^210^{\log g-\log g_\odot},
\end{equation}

with $\log g_\odot=4.437$. The same stellar radius is used when calculating the transit depth through $(R_{\rm p}/R_\star)^2$.

The Earth-equivalent semi-major axis is then

\begin{equation}
\frac{a}{\mathrm{AU}}=\sqrt{\frac{L_\star}{L_\odot}},
\label{eq:semi_major_axis}
\end{equation}

and the orbital period follows from Kepler's third law,

\begin{equation}
\left(\frac{P}{1\,{\rm yr}}\right)^2=\frac{(a/{\rm AU})^3}{M_\star/M_\odot}.
\label{eq:orbital_period}
\end{equation}

This recipe provides a common Earth-twin geometry across the stellar grid rather than reproducing the orbital properties of any known planet, and is consistent in spirit with placing the planet at Earth-like irradiation within the classical habitable-zone framework \citep{Kopparapu2013}. The orbital configuration and the epoch-dependent irradiation used in the atmospheric calculations are treated separately. For each stellar proxy, the semi-major axis is fixed using its present-day luminosity and is kept unchanged across all four geological epochs. The luminosity scaling factors $f_L$ obtained from the Earth-through-time framework \citep{Claire2012,Rugheimer2018} are instead applied to the stellar spectrum supplied to the photochemical calculation. Thus, the epoch-dependent scaling modifies the irradiation driving the atmospheric chemistry and thermal structure, while the orbital distance, period, and transit geometry remain fixed for a given host star.

We assume a fixed impact parameter of $b=0.3$ for all systems. For a circular orbit, the transit duration of a planet is calculated following \citet{Winn2010} as

\begin{equation}
T_{14}=\frac{P}{\pi}\arcsin\left[\frac{R_\star}{a}\sqrt{\frac{(1+R_{\rm p}/R_\star)^2-b^2}{1-(bR_\star/a)^2}}\right].
\label{eq:t14}
\end{equation}

Each simulated transit consists of an in-transit interval of duration $T_{14}$ together with an out-of-transit baseline of equal duration. In theory, this is a conservative estimate, since partial-transit observations with sufficient out-of-transit baseline may also yield a detection. The effective cadence is

\begin{equation}
t_{\rm cad}=t_{\rm exp}+t_{\rm duty},
\end{equation}

where $t_{\rm duty}=34$\,s. The number of spectra obtained per transit event is therefore

\begin{equation}
N_{\rm per\,transit}=\frac{2T_{14}}{t_{\rm cad}},
\label{eq:npertransit}
\end{equation}

with $T_{14}$ and $t_{\rm cad}$ expressed in same units. The required number of transit observations is then

\begin{equation}
N_{\rm transits}=\frac{N_{\rm frames}}{N_{\rm per\,transit}}
\label{eq:ntransits}
\end{equation}

To facilitate comparison between systems with substantially different transit durations, we additionally express the observing requirement in equivalent 7\,h units of on-sky telescope time following \cite{KurzawaFerrandez2026},

\begin{equation}
N_{\rm nights}=\left[\frac{2N_{\rm transits}T_{14}}{7\,{\rm h}}\right]
\label{eq:nnights}
\end{equation}

Here, $N_{\rm nights}$ represents the total on-sky observing time expressed in 7\,h units and should not be interpreted as the number of calendar nights required to complete the observations. Actual scheduling feasibility will additionally depend on transit ephemerides, seasonal visibility, weather losses, and other operational constraints. Table \ref{tab:host_observing} summarizes the stellar and orbital properties used for the detectability calculations. The distance to each stellar proxy determines the photon budget, while the adopted stellar radius sets the transmission contrast. 

\section{Results}
\label{sec:results}

\subsection{Photochemical evolution from Archean to modern conditions}
\label{sec:results:chemistry}

The Archean calculations show the expected transition from weakly reducing to methane-rich, haze-forming chemistry (see Figure \ref{fig:archean_chemisrty}). N$_2$ is treated as the dominant background atmospheric constituent and is therefore not shown in the figure, while the imposed CH$_4$/CO$_2$ ratio controls the reduced-carbon inventory. Increasing CH$_4$/CO$_2$ raises the CH$_4$ abundance throughout the column and enhances secondary hydrocarbons, including C$_2$H$_6$ and haze precursors. The strongest haze production occurs in the high-CH$_4$/CO$_2$ members of each climate family, broadly consistent with the behavior identified by \citet{Arney2016} and laboratory haze experiments \citep{Trainer2006}. O$_2$ remains fixed near the anoxic lower-boundary value and O$_3$ is negligible. Thus, within this grid, methane and its photochemical products provide the dominant spectroscopic distinction among the Archean models.

\begin{figure*}
\centering
\begin{minipage}[b]{0.49\textwidth}\includegraphics[width=\textwidth]{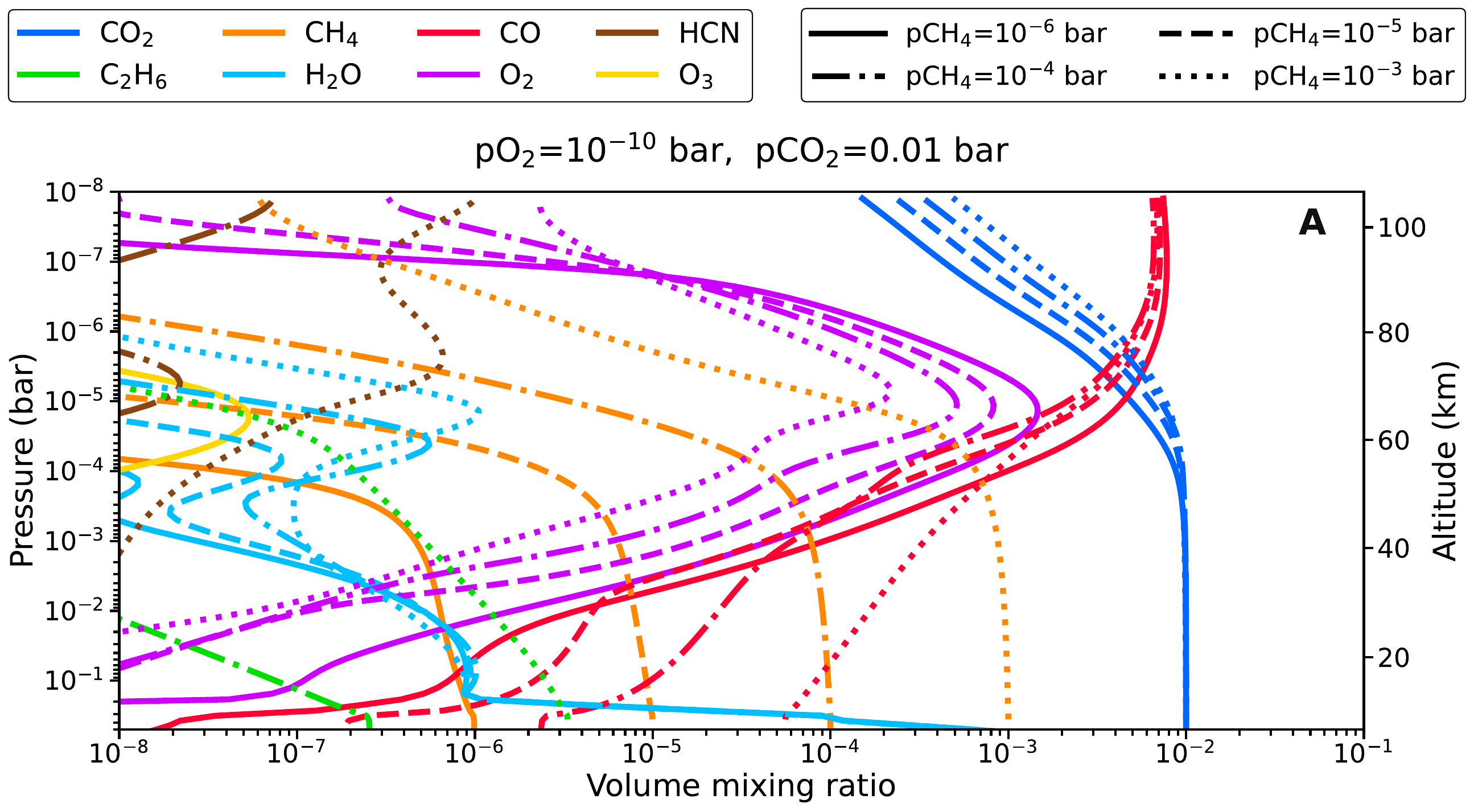}\end{minipage}
\begin{minipage}[b]{0.49\textwidth}\includegraphics[width=\textwidth]{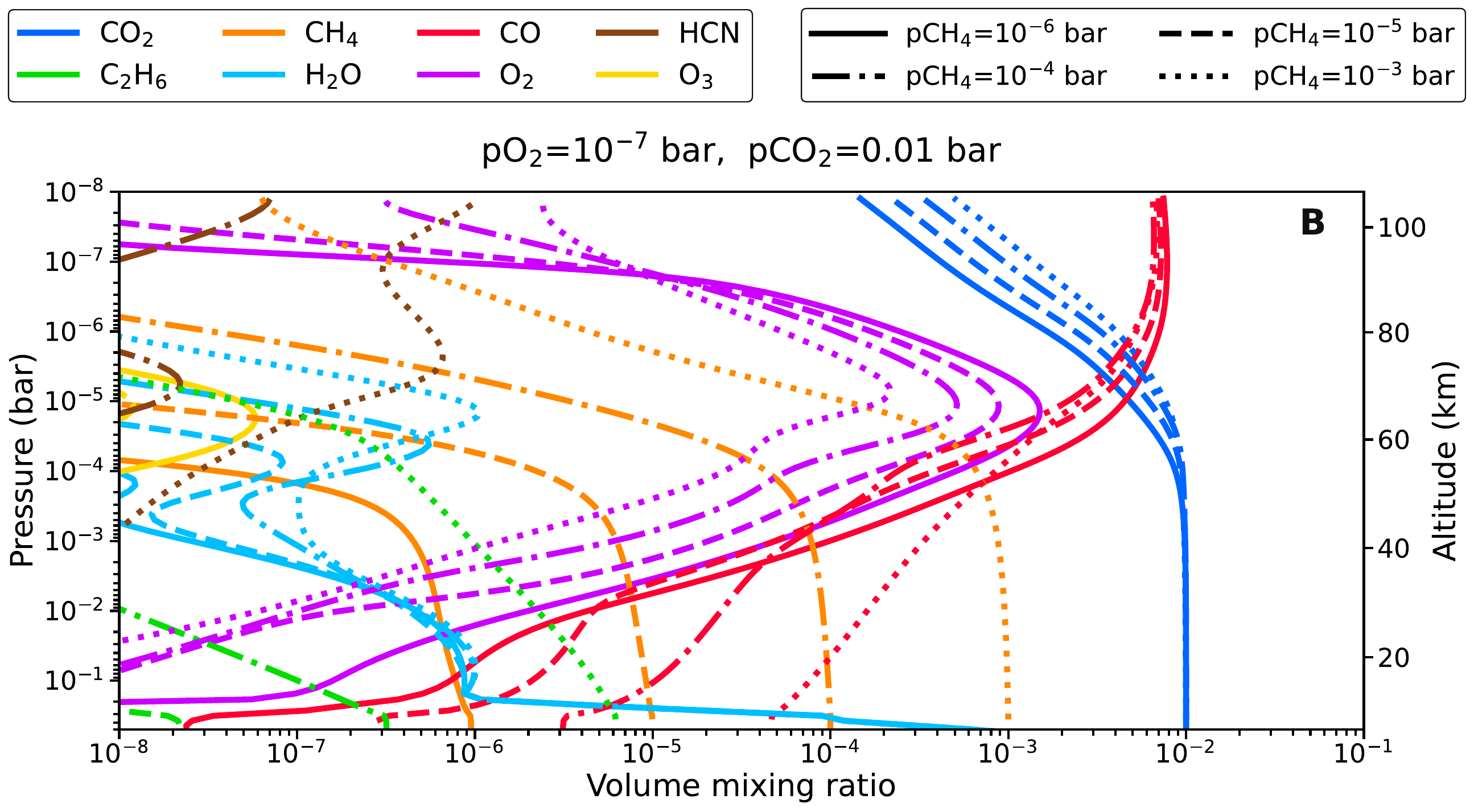}\end{minipage}\\[0.6cm]
\begin{minipage}[b]{0.49\textwidth}\includegraphics[width=\textwidth]{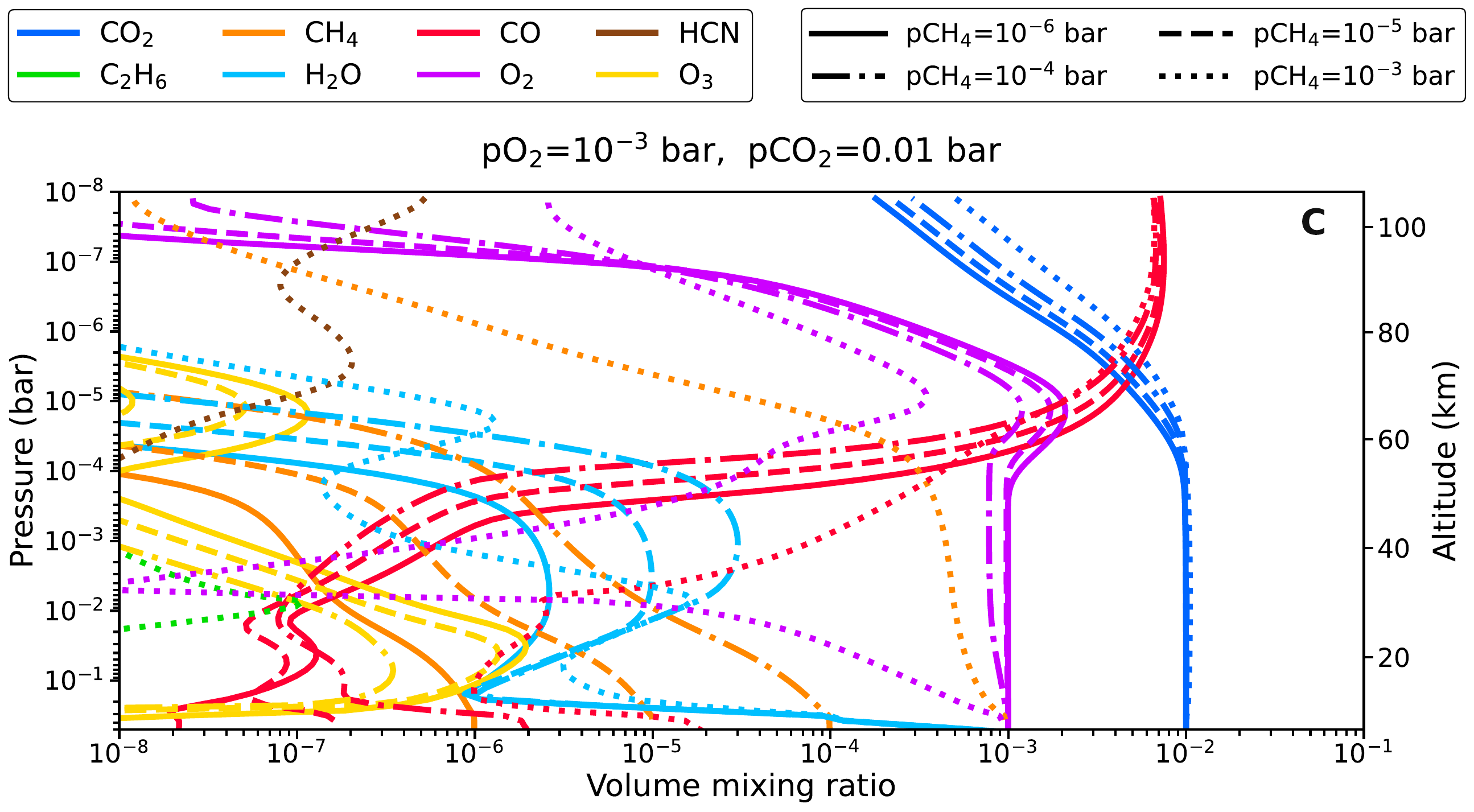}\end{minipage}
\caption{Steady-state mixing-ratio profiles for the Proterozoic grid at surface O$_2$ mixing ratios of $10^{-10}$, $10^{-7}$, and $10^{-3}$. Within each panel, CH$_4$ increases from $10^{-6}$ to $10^{-3}$ at the lower boundary.}
\label{fig:proterozoic_chemisrty}
\end{figure*}

The Proterozoic grid separates the effects of atmospheric oxidation and methane abundance (see Figure \ref{fig:proterozoic_chemisrty}). At $p{\mathrm{O}_2}=10^{-10}$ bar, the column remains strongly anoxic and the CH$_4$ profiles largely track the imposed lower boundary. Raising O$_2$ to $10^{-7}$ bar produces an intermediate redox state in which substantial reduced carbon can coexist with a larger oxidizing reservoir. At $10^{-3}$ bar, O$_2$ becomes a major constituent and an O$_3$ layer develops, particularly in the methane-poor models.

\begin{figure}
\centering
\includegraphics[width=\columnwidth]{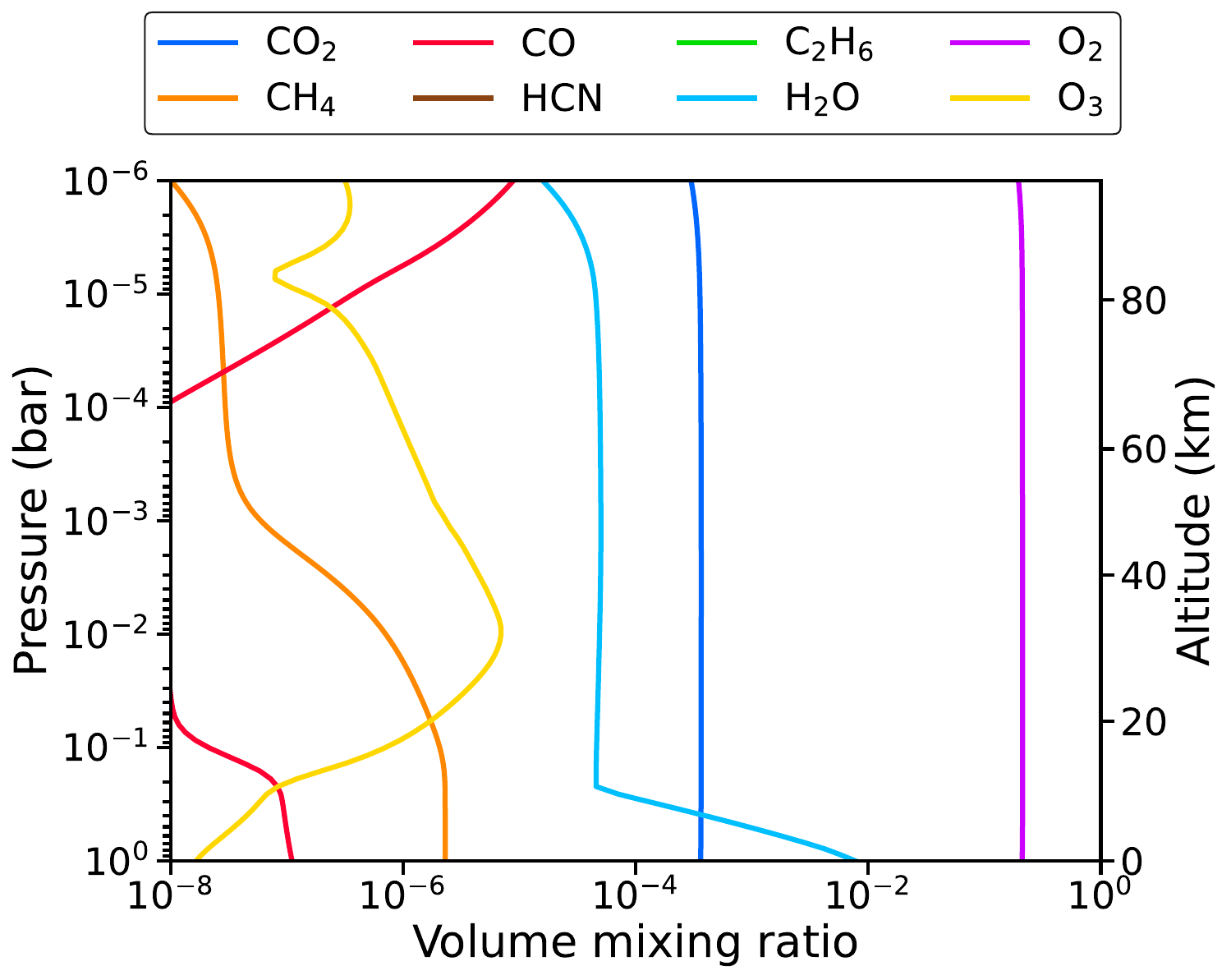}
\caption{Steady-state mixing-ratio structure of the modern Earth benchmark. The atmosphere is O$_2$ rich, contains a stratospheric O$_3$ layer, and retains only trace CH$_4$ relative to the Archean and Proterozoic methane-rich cases.}
\label{fig:modern_chemisrty}
\end{figure}

The modern Earth benchmark (Figure \ref{fig:modern_chemisrty}) is qualitatively distinct from both earlier grids: O$_2$ dominates the oxidizing inventory, the stratospheric O$_3$ layer is well developed, and CH$_4$ is strongly suppressed relative to the high-methane Archean and Proterozoic cases. This atmospheric state provides the conventional reference for comparison with the earlier Archean- and Proterozoic-like stages of Earth’s atmospheric evolution considered here.

\begin{figure*}
\centering
\includegraphics[width=\textwidth]{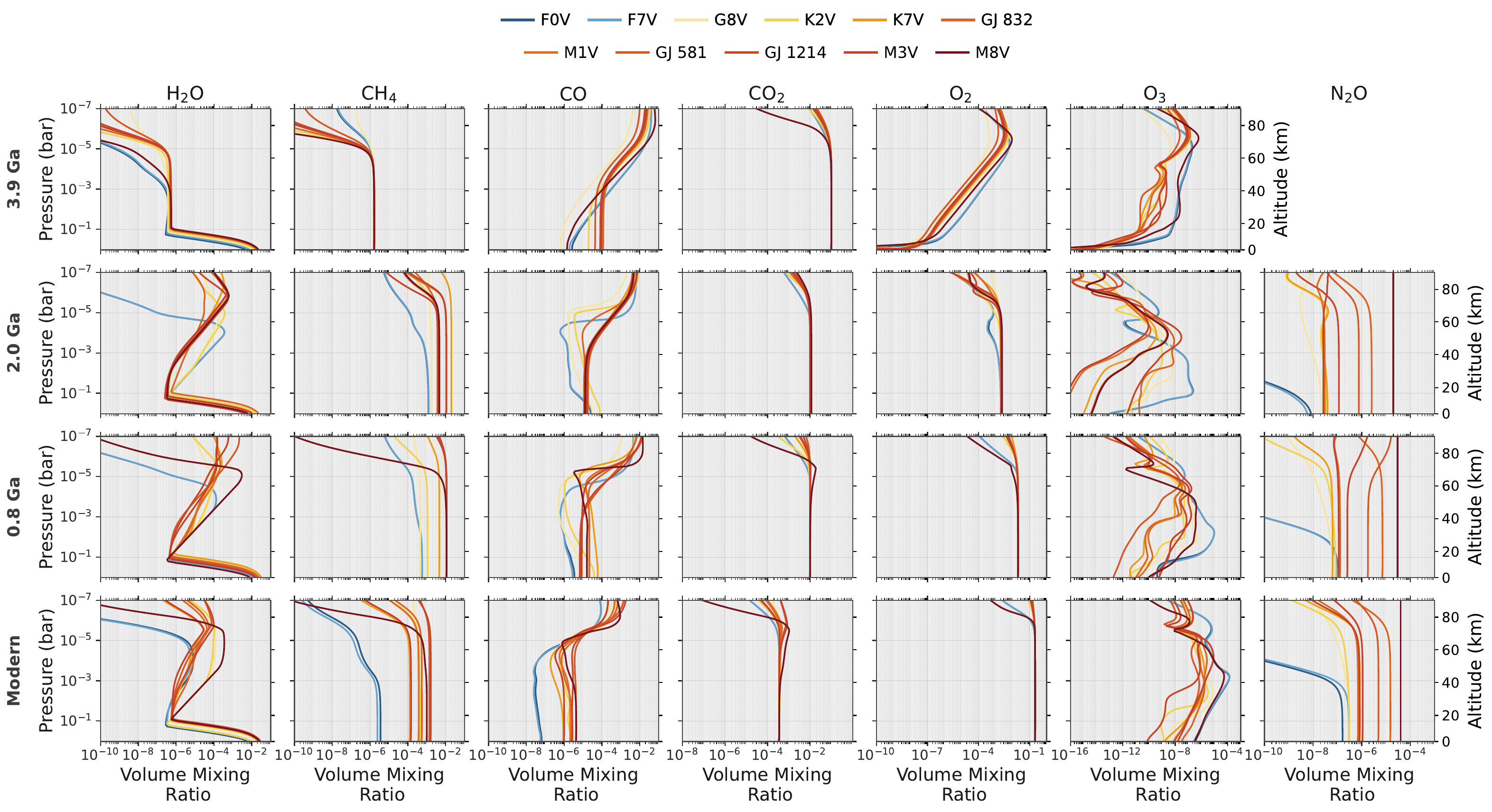}
\caption{Representative mixing-ratio profiles for the FGKM host $\times$ geological-epoch grid following \citet{Rugheimer2018}. The dominant temporal trend is the rise of O$_2$ and O$_3$ with geological age, while host-star UV irradiation alters the abundance and vertical distribution of photochemically sensitive species such as CH$_4$.}
\label{fig:grid_VMR}
\end{figure*}

The multi-star grid retains the epoch ordering imposed by the lower-boundary conditions while introducing a strong host-star dependence (the mixing ratio profiles for all species across stellar hosts are shown in Figure \ref{fig:grid_VMR}). The 3.9\,Ga atmospheres are effectively anoxic for all hosts. Oxygen increases at 2.0\,Ga and 0.8\,Ga and reaches the modern value in the final epoch. At fixed oxygenated epoch, CH$_4$ is generally enhanced around cooler hosts, consistent with previous calculations showing that the UV environments of K and M dwarfs can lengthen the atmospheric lifetime of reduced species \citep{segura2005biosignatures,Rugheimer2015,Rugheimer2018}. The host spectrum therefore modulates the detailed chemical state without erasing the broad evolution from anoxic to oxic conditions.

\subsection{High-resolution spectral evolution}
\label{sec:results:spectra}

\begin{figure*}
\centering
\includegraphics[width=0.9\textwidth]{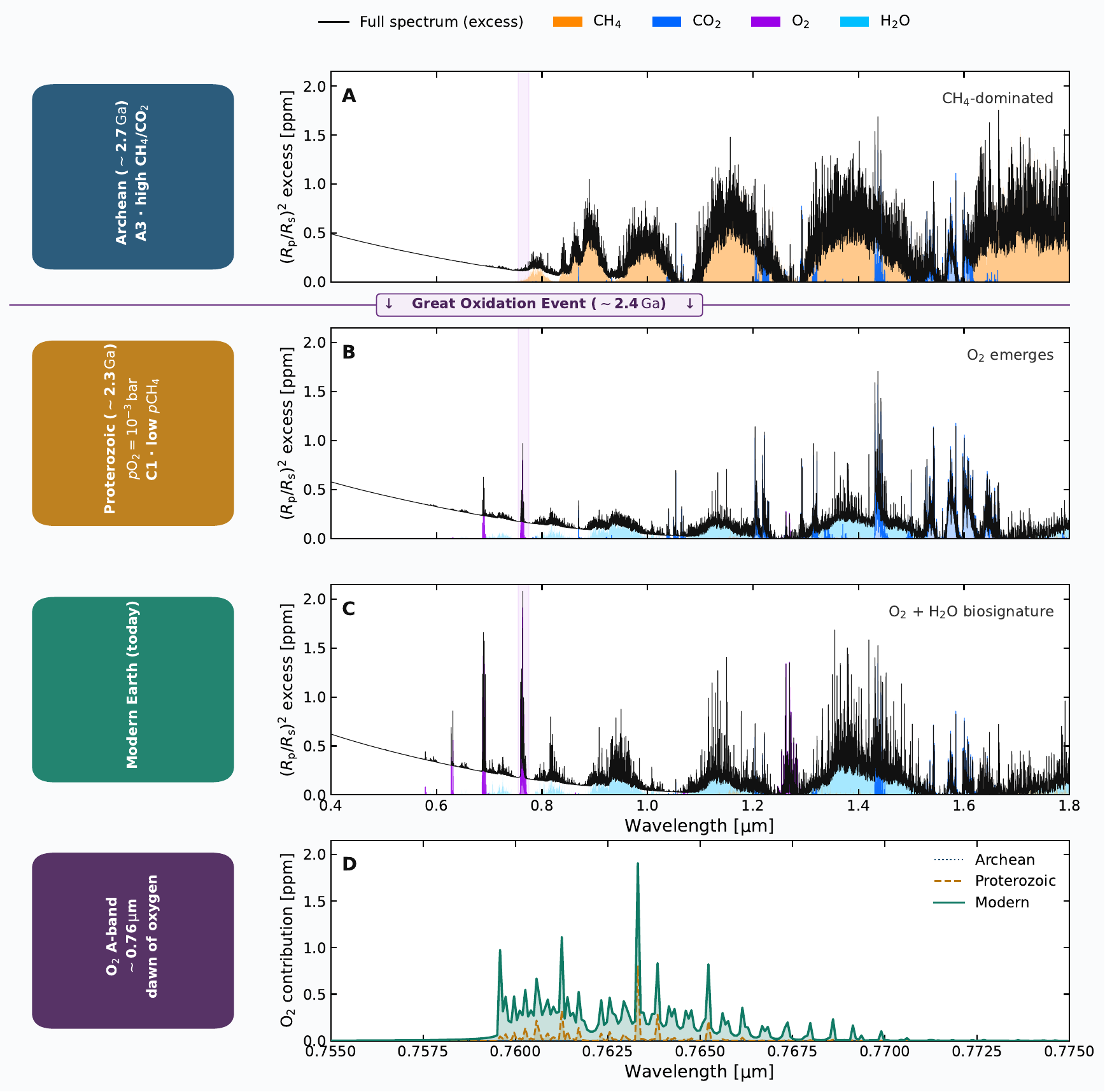}
\caption{Clear-atmosphere transmission spectra at $R\sim10^5$ for representative Archean (A3-H1), Proterozoic (C1-V2), and modern Earth cases. Molecular contributions are shown over 0.4--1.8\,$\mu$m, with the lower panel emphasizing the growth of the O$_2$ A-band as the atmosphere becomes oxygenated. This optical O$_2$ feature lies outside the YJH wavelength range used for the ANDES detectability calculations.}
\label{fig:time_evolution_spectra}
\end{figure*}

Figure \ref{fig:time_evolution_spectra} summarizes the principal high-resolution spectral changes for the Sun-Earth system over geological timescale. The Archean spectrum is methane dominated, with small additional features from H$_2$O and CO$_2$. The selected Proterozoic model develops strong O$_2$ absorption while retaining a chemically distinct reduced component. Modern Earth produces the deepest O$_2$ features and a dense H$_2$O line forest, whereas CH$_4$ is comparatively weak. The O$_2$ A-band near 0.76\,$\mu$m provides the clearest visual tracer of the transition from anoxic to oxic conditions in the full optical-near-infrared spectrum.

The detailed Archean and Proterozoic spectral grids are shown in Appendix Figures \ref{fig:archean_spectra} and \ref{fig:proterozoic_spectra}. Across the Archean H1 families, increasing CH$_4$/CO$_2$ strengthens methane features throughout the bandpass. In the Proterozoic grid, two things are noticeable: increasing O$_2$ introduces the A-band and other weaker O$_2$ features, while increasing CH$_4$ strengthens the methane bands. These trends demonstrate that a single epoch label does not correspond to a unique spectrum; the observable state depends on the redox boundary conditions within that epoch.

Appendix Figure \ref{fig:modern_spectra} provides a molecule-by-molecule decomposition of the modern spectrum. O$_2$ and H$_2$O dominate the strongest optical and near-infrared features in our wavelength range while CO$_2$ and CH$_4$ contribute weakly. This decomposition motivates the four molecules carried into the detectability analysis: O$_2$ and CH$_4$ as redox-sensitive tracers, and H$_2$O and CO$_2$ as environmental and climate context.

\subsection{ANDES detectability}
\label{sec:results:detectability}

\begin{figure}
\centering
\includegraphics[width=\columnwidth]{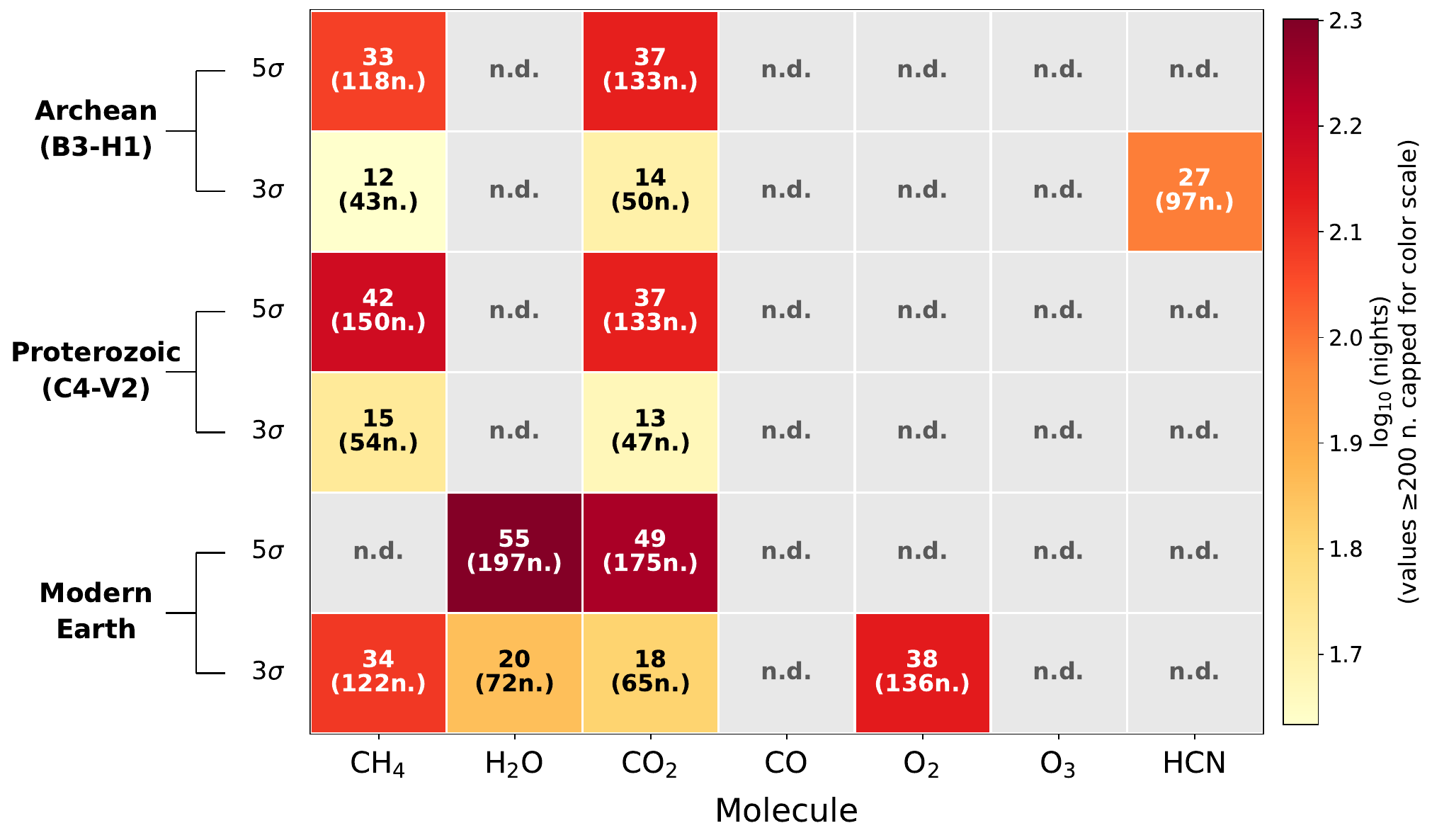}
\caption{Estimated 5$\sigma$ and 3$\sigma$ detectability requirements in terms of transits ($N_{\mathrm{transits}}$) and equivalent observing nights ($N_{\mathrm{nights}}$) for representative solar-host Archean, Proterozoic, and modern-Earth templates using ANDES YJH observations. Here, $\sigma$ refers to the likelihood-based cross-correlation significance defined in Section \ref{sec:ratri}. The color scale shows $\log_{10}(N_{\mathrm{nights}})$ for configurations within the observing limits. Cell annotations indicate $N_{\mathrm{transits}}$ and the corresponding number of equivalent 7\,h observing nights. Configurations requiring more than 500 transits or 200 equivalent observing nights are shown as gray cells and marked n.d.}
\label{fig:5sigma_modern_earth_detec}
\end{figure}

Figure \ref{fig:5sigma_modern_earth_detec} shows the observing requirements for reaching 3$\sigma$ and 5$\sigma$ detections for the representative solar-twin atmospheres. The methane-rich Archean model is most favorable for CH$_4$, requiring 33 transits to reach 5$\sigma$, while O$_2$ remains effectively inaccessible because of the anoxic atmospheric composition. In the selected Proterozoic case, CH$_4$ reaches 5$\sigma$ after 42 transits, whereas the O$_2$ requirement exceeds 500 transits (200 equivalent observing nights) even at 3$\sigma$. Modern Earth produces the strongest O$_2$ signal of the three representative cases and reaches 3$\sigma$ after 38 transits, although a 5$\sigma$ detection still exceeds the required limits in the YJH-only configuration. It illustrates that atmospheric oxygenation strengthens the O$_2$ signal substantially, but its detectability remains limited by the intrinsically small scale height of a high-mean-molecular-weight terrestrial atmosphere and by the exclusion of the strong 0.76\,$\mu$m O$_2$ A-band from the YJH-only configuration.

\begin{figure*}
\centering
\includegraphics[width=\textwidth]{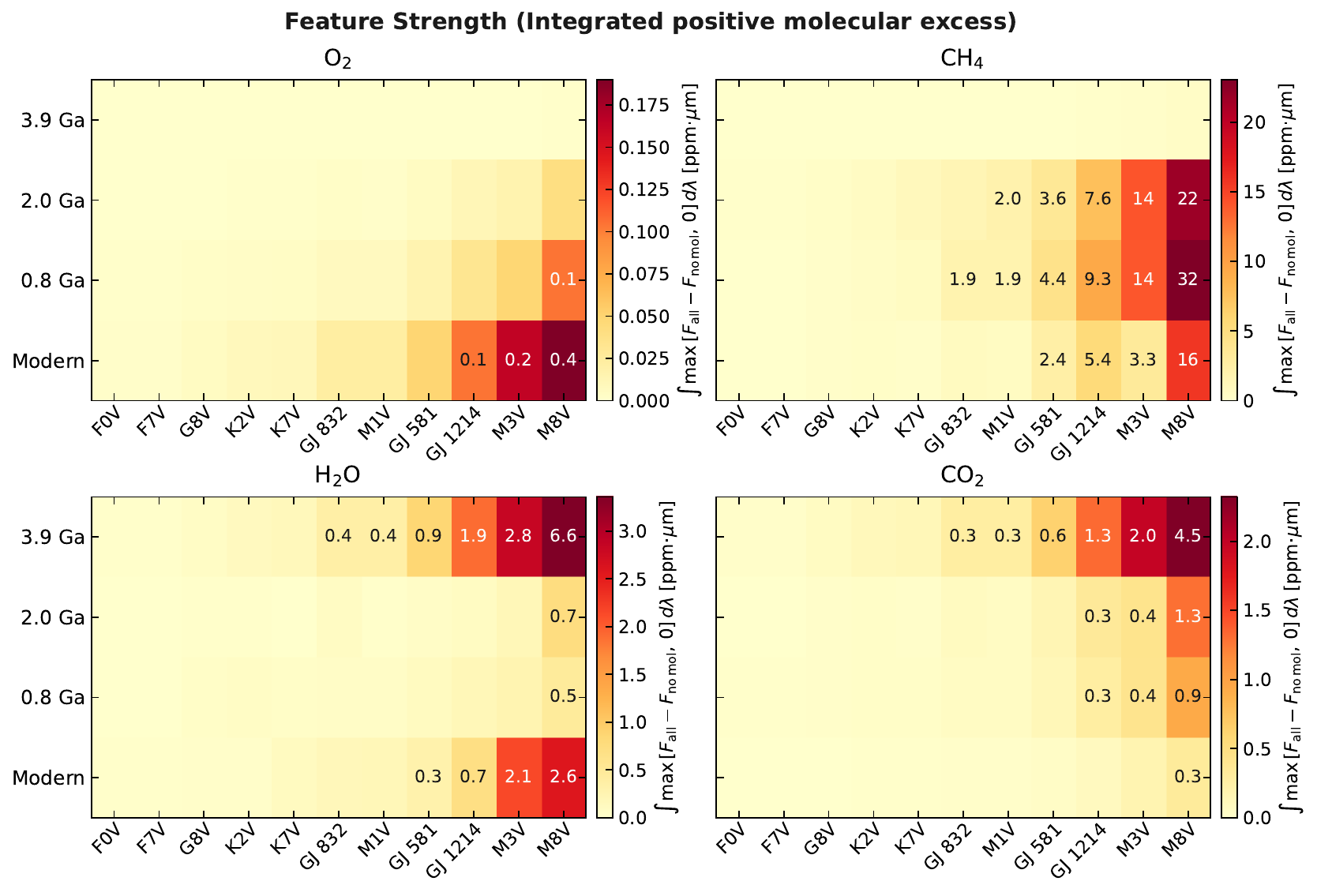}
\caption{Feature-strength heatmaps for the FGKM hosts $\times$ epoch grid for O$_2$, CH$_4$, H$_2$O, and CO$_2$. Columns represent different hosts (hot F to cool M) and rows are geological epochs (3.9\,Ga, 2.0\,Ga, 0.8\,Ga, modern). Color shows the wavelength-integrated positive leave-one-out transit-depth excess over $0.4$--$1.8\,\mu$m. CH$_4$ dominates the integrated budget for cooler hosts, while O$_2$ remains negligible at 3.9\,Ga and strengthens mainly for late-M modern cases.}
\label{fig:heatmap_diff_host}
\end{figure*}

\begin{figure*}
\centering
\begin{minipage}[b]{0.49\textwidth}\includegraphics[width=\textwidth]{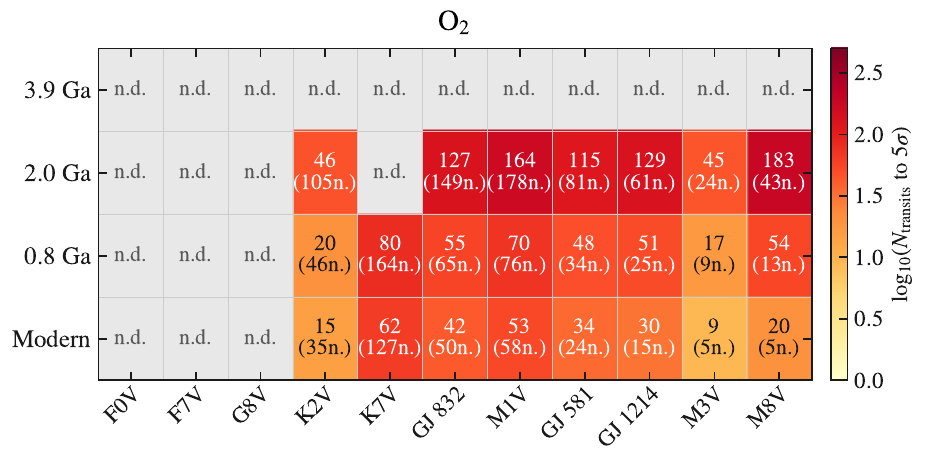}\end{minipage}
\begin{minipage}[b]{0.49\textwidth}\includegraphics[width=\textwidth]{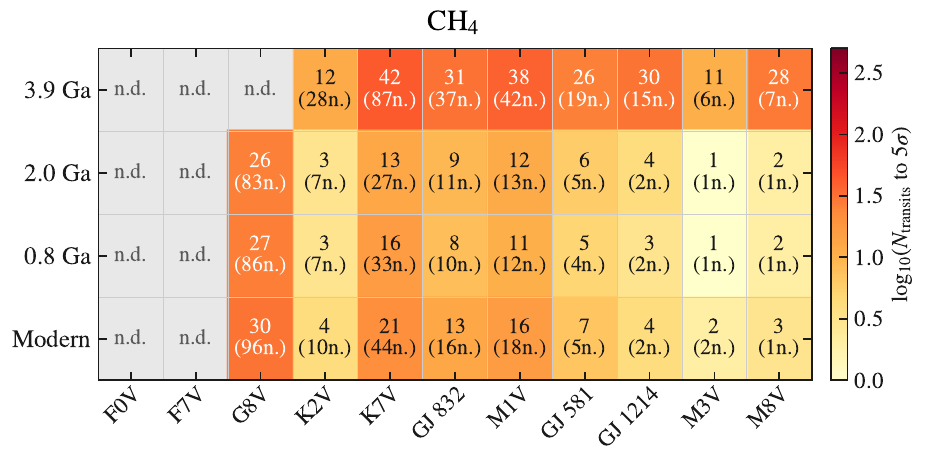}\end{minipage}
\begin{minipage}[b]{0.49\textwidth}\includegraphics[width=\textwidth]{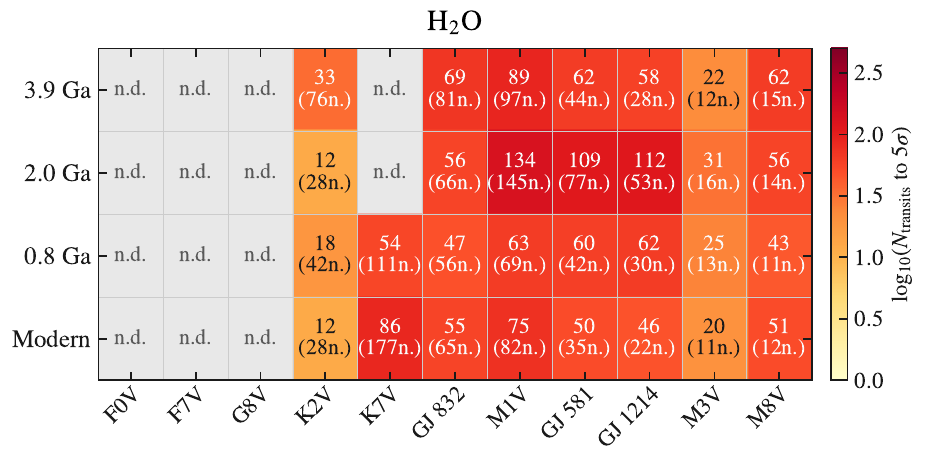}\end{minipage}
\begin{minipage}[b]{0.49\textwidth}\includegraphics[width=\textwidth]{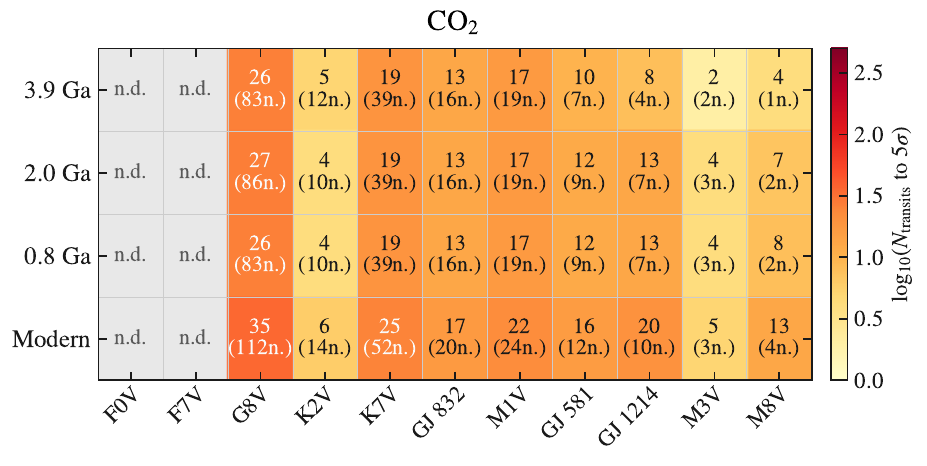}\end{minipage}
\caption{$5\sigma$ requirement heatmaps for O$_2$, CH$_4$, H$_2$O, and CO$_2$ across FGKM hosts using ANDES YJH-band observations. Here, $\sigma$ refers to the likelihood-based cross-correlation significance defined in Section \ref{sec:ratri}. The color scale shows $\log_{10}(N_{\mathrm{transits}})$ for configurations within the observing limits. Cell annotations give $N_{\mathrm{transits}}$ and the corresponding number of equivalent 7\,h observing nights. Configurations requiring more than 500 transits or 200 equivalent observing nights are shown as gray cells and marked n.d.}
\label{fig:5sigma_detectability_diff_host}
\end{figure*}

Figure \ref{fig:heatmap_diff_host} extends the spectral comparison across the full FGKM host-star grid. The heatmaps show the wavelength-integrated leave-one-out contribution of O$_2$, CH$_4$, H$_2$O, and CO$_2$ for each geological epoch. O$_2$ contributes negligibly to the 3.9\,Ga spectra for all hosts, as expected for the anoxic atmosphere, and its contribution increases toward the oxygenated epochs. CH$_4$ shows a different pattern, with strong features persisting for several of the cooler stellar hosts even after atmospheric oxygenation. Hence, the molecular spectrum changes with both geological epoch and stellar host, reflecting the host-dependent atmospheric abundances produced by the photochemical models.

Figure \ref{fig:5sigma_detectability_diff_host} shows how these spectral differences translate into the number of transits and equivalent observing nights required for a 5$\sigma$ detection with ANDES in the YJH bands. O$_2$ does not reach a 5$\sigma$ detection within the observing limits for the 3.9\,Ga atmospheres and becomes gradually more detectable toward modern Earth. For modern Earth, the lowest equivalent observing-time requirements for O$_2$ occur for the M3V and M8V proxies, requiring 9 and 20 transits, respectively, corresponding to approximately 5 equivalent observing nights in both cases. CH$_4$ and CO$_2$ are generally easier to detect around the cooler hosts; for the modern M3V case they require only 2 and 5 transits, respectively, compared with 20 transits for H$_2$O. H$_2$O also remains more demanding than CH$_4$ and CO$_2$ for the M8V proxy, requiring 51 transits in the modern case. The F0V and F7V cases remain outside the observing limits for all four molecules, while O$_2$ and H$_2$O also remain non-detectable for the G8V star.

\begin{figure}
\centering
\includegraphics[width=\columnwidth]{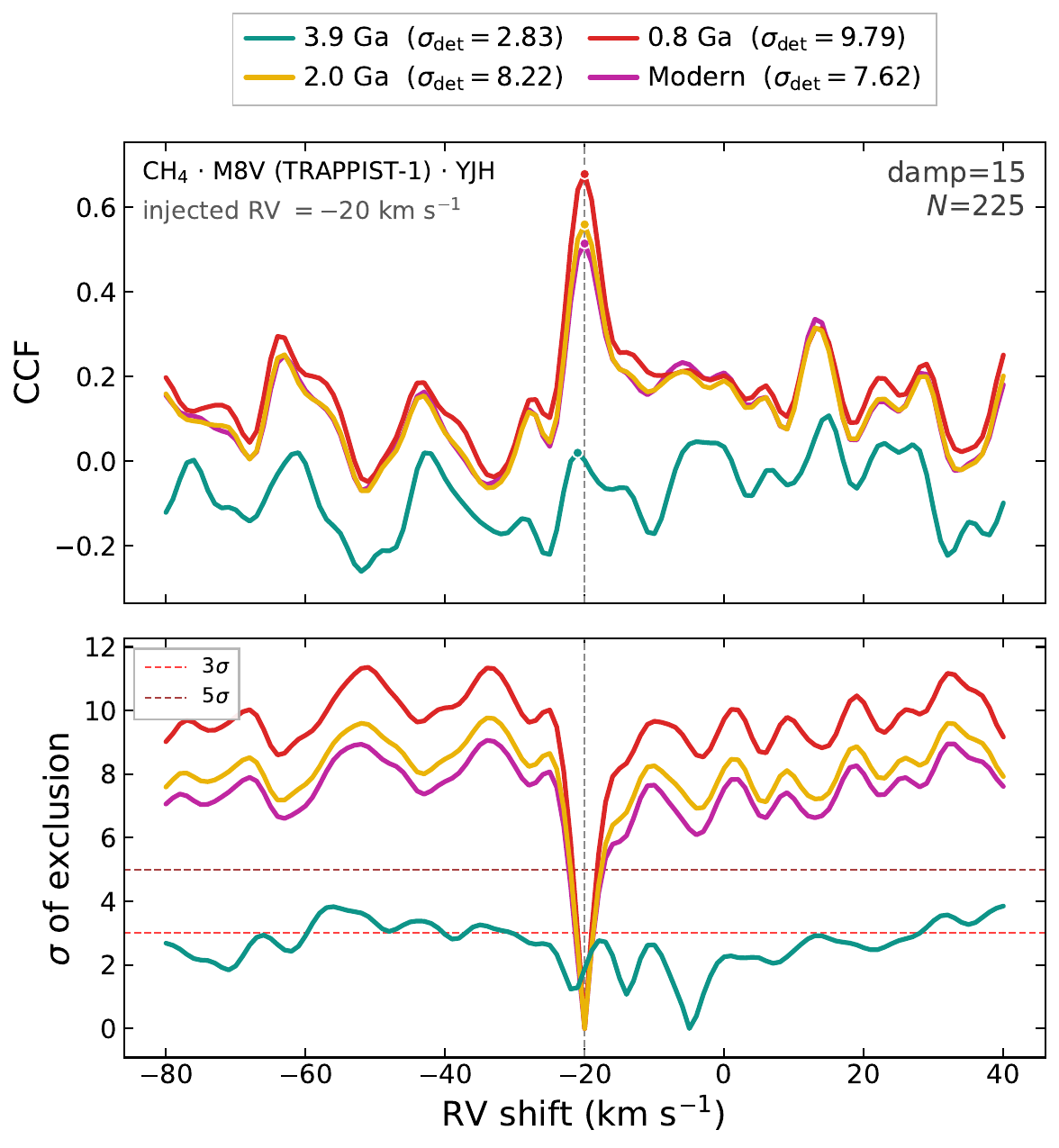}
\caption{Stacked CH$_4$ CCFs for the M8V (TRAPPIST-1) proxy at the four geological epochs using 225 frames ($t{\mathrm{exp}}\approx90.5$ s, airmass $=1.2$). (Top): CCF versus RV shift; the dashed vertical line marks the time-invariant injected systemic radial velocity ($-20$\,km\,s$^{-1}$). (Bottom): significance with which each trial RV can be statistically excluded as the planet's true velocity; the curve dips sharply to its minimum at the injected RV with $3\sigma$ and $5\sigma$ thresholds marked for reference (dashed horizontal lines). The depth of this dip relative to the surrounding curve gives the detection significance for each epoch: 3.9\,Ga (2.83$\sigma$ - a non-detection), 2.0\,Ga (8.22$\sigma$), 0.8\,Ga (9.79$\sigma$), modern (7.62$\sigma$).}
\label{fig:ccf_m8v_ch4}
\end{figure}

Figure \ref{fig:ccf_m8v_ch4} provides one example of the cross-correlation signal underlying these detectability estimates. For the M8V host, the CH$_4$ CCF changes substantially between geological epochs even though the observing setup remains unchanged. In the 225-frame example, the 3.9\,Ga atmosphere remains below 3$\sigma$, whereas the 2.0\,Ga, 0.8\,Ga, and modern cases exceed 5$\sigma$. With the photon-noise scaling used for the detectability calculations, the 3.9\,Ga case requires 28 transits to reach the 5$\sigma$ criterion, compared with only 2--3 transits for the oxygenated epochs. This sequence follows the CH$_4$ abundances obtained for the M8V photochemical models and illustrates why strong methane absorption is not restricted to the Archean atmosphere in the host-star grid.

\section{Discussion}
\label{sec:discussion}

\subsection{What is required to distinguish geological states?}
\label{sec:discussion:era}

The main result of our calculations indicates that geological state cannot be inferred from a single molecule. Methane provides the clearest example. The solar-host Archean grid is deliberately methane rich and produces strong CH$_4$ signatures, whereas the 3.9\,Ga models in the \citet{Rugheimer2018} grid use much lower CH$_4$ boundary abundances. At the same time, cooler host stars can preserve substantial methane after oxygenation. A strong CH$_4$ detection is compatible with both an anoxic, methane-rich atmosphere and an oxygenated atmosphere around a UV environment that permits CH$_4$ to accumulate \citep{segura2005biosignatures,Rugheimer2015}.

O$_2$ provides the principal discriminant between these cases, but only if the observations are sensitive to the relevant oxygenation level. In our Proterozoic grid, the $p\mathrm{O}_2=10^{-7}$ bar cases are spectroscopically much closer to the anoxic states than to the $10^{-3}$ bar cases. A non-detection of O$_2$ thus constrains geological state only when the achieved upper limit lies below the O$_2$ abundance expected for the intermediate atmosphere being tested. For this reason, reporting the sensitivity reached for O$_2$ is as important as reporting a formal detection.

CO$_2$ and H$_2$O provide complementary environmental information. In the Earth-through-time grid, their abundances are tied to the implemented evolutionary boundary conditions, but on real planets neither molecule maps uniquely onto age. CO$_2$ is governed by outgassing, weathering, and climate feedback, while H$_2$O depends on the thermal structure and water inventory. Their role is primarily contextual, i.e., they help determine whether the O$_2$-CH$_4$ combination is physically plausible for the inferred climate state.

This distinction is also important for biosignature interpretation. An atmosphere classified as oxygenated is not automatically biological, and a geological-era analogue is not by itself a biosignature. Abiotic O$_2$ production is especially relevant for M-dwarf planets \citep{Meadows2018}. Any interpretation of joint O$_2$ and CH$_4$ therefore requires the broader atmospheric and stellar context, consistent with chemical-disequilibrium arguments \citep{KrissansenTotton2018,KrissansenTotton2020}.

\subsection{Relation to previous Earth-through-time and ANDES studies}
\label{sec:discussion:comparison}

Our atmospheric grid is directly motivated by \citet{Zahnle2006}, \citet{Arney2016}, and \citet{Rugheimer2018}, who calculated self-consistent climate and photochemistry for Earth-like planets across four geological epochs and an FGKM stellar grid. Their study emphasized the low- to moderate-resolution spectral consequences of geological evolution. We retain the same conceptual structure but recompute the chemistry with \texttt{Photochem}, use modern MUSCLES UV spectra as host proxies, and propagate the resulting columns as inputs into $R\sim10^5$ transmission spectra and ANDES cross-correlation studies. Differences in chemical network, fixed temperature structure, stellar UV inputs, and aerosol treatment prevent a one-to-one profile comparison.

The present study is complementary to the modern-Earth ANDES analysis of \citet{KurzawaFerrandez2026} in terms of detectability, but with the evolution of atmospheric chemistry more akin to \citet{Currie2023} (but uses a larger grid of stellar proxies by including stars other than M dwarfs). \citet{KurzawaFerrandez2026} evaluated the detectabilities of H$_2$O, CO$_2$, CH$_4$, and O$_2$ for 18 known potentially habitable transiting planets, but by assuming the same modern-Earth atmosphere for all cases. Here, we instead vary both the geological state and host-star spectrum for idealized Earth twins. This means that our modern-Earth analogue is not similar to the modern-Earth modeled in \citet{KurzawaFerrandez2026}, but instead a context dependent scenario based on both the geological epoch and the star. While both approaches identify cool M-dwarf systems as the most favorable transmission targets, our detectability time estimates for H$_2$O are much higher compared to \citet{KurzawaFerrandez2026}'s estimates because CH$_4$ starts dominating our modeled spectra for modern Earth atmospheres around such stars compared to always dominant H$_2$O for their study. This discrepancy between H$_2$O and CH$_4$ is also seen in \citet{Currie2023}, who also use an epoch and host star dependent evolving atmospheric chemistry setup. They also find a trend of CH$_4$ and CO$_2$ being substantially easier to detect in comparison to O$_2$ and H$_2$O for modern Earth like analogues (labeled as pre-industrial Earth for their study) around M dwarf stars, akin to our study.

Our most favorable modern-Earth O$_2$ cases require approximately 5 equivalent observing nights or 9 and 20 transits to reach a 5$\sigma$ detection threshold for the M3V and M8V hosts respectively, placing them towards the optimistic end of previous high-resolution studies. Earlier studies typically found requirements of tens to hundreds of transits for comparable O$_2$ searches \citep{Snellen2013,RodlerLopezMorales2014,SerindagSnellen2019,Currie2023} depending on the complexity of the simulations arising from whether atmospheric refraction was included in spectrum generation or a moving exoplanet signal was injected or more realistic treatment of white and red noise was included, as well as the way a molecular detection was quantified (which differs across most papers). More recently, \citet{deming2025minimizing} showed that stellar heterogeneity can further complicate such observations: for a temperate rocky planet around an M4V star at 12\,pc, their photon-noise-limited simulation reached only $S/N \simeq 4$ for O$_2$ after 200 transits, with unocculted star spots introducing an additional source of correlated noise. Direct comparison between these studies is, thus, difficult because the specific targets, wavelength ranges, noise prescriptions, and detection metrics are not uniform across them. In our case, the relatively short observing requirements are driven by the use of very nearby stellar proxies and by the assumption that the signal continues to improve according to photon-noise-limited $\sqrt{N}$ scaling after idealized removal of telluric and stellar contamination. Hence, the resulting observing times should be interpreted as optimistic estimates for the specific proxy systems considered here.

The YJH band restriction for ANDES as assumed here is especially relevant for O$_2$. The 0.76\,$\mu$m A-band, which dominates the optical comparison in Figure \ref{fig:time_evolution_spectra}, is not included in the detectability maps. The YJH analysis presented here relies primarily on weaker near-infrared O$_2$ features, including the 1.27\,$\mu$m band, which is itself affected by terrestrial O$_2$ and airglow. Extending the same framework to the optical ANDES arms will be essential for a complete assessment of oxygen detectability.

\subsection{Host-star dependence and target selection}
\label{sec:discussion:hosts}

Our choice for cool hosts in the detectability maps is primarily motivated by two reasons. The first is geometric. For a fixed terrestrial atmosphere, the approximate amplitude of a transmission feature scales as $R_{\rm p}H/R_\star^2$, making the same Earth twin substantially easier to observe around an M dwarf than around an F or G star. In our proxy grid, difference in contrast can vary over two orders of magnitude between the 1.57\,$R_\odot$ WASP-17 proxy and the 0.12\,$R_\odot$ TRAPPIST-1 proxy. Longer transits around larger stars do not compensate for that loss.

The second contribution arises from the stellar radiation environment. The host UV spectrum controls photolysis and therefore the steady-state abundances of molecules such as CH$_4$, O$_3$, and N$_2$O. Consequently, a favorable stellar radius does not merely increase the transmission contrast; the stellar spectrum can also change the atmosphere being observed. This coupling is why a host-star grid is necessary when discussing geological analogues. The stellar proxies represent specific stars, and their distances affect the exposure times required to reach a given signal-to-noise ratio. Very nearby systems such as $\varepsilon$ Eri and GJ\,832 can saturate the CCDs even at the minimum exposure time, whereas more distant analogues would require longer integrations to achieve the same S/N per pixel. 

\subsection{Model limitations and observational implications}
\label{sec:discussion:limitations}

Several assumptions make the present analysis optimistic. First, clouds and aerosol extinction are omitted from the transmission calculation. This is particularly important for the Archean models, for which organic haze is chemically active and would reduce the short-wavelength continuum available for transmission spectroscopy \citep{Arney2016}. The reported Archean detectability estimates reflect as a cloud-free upper limit for our study. Second, the atmospheric temperature structures are assumed to be adiabatic profiles and are not iterated self-consistently with the atmospheric chemistry. The photochemical calculations are also one-dimensional and use constant vertical mixing and a single representative stellar zenith angle, thereby neglecting three-dimensional circulation and morning-evening terminator heterogeneity. Third, the list of opacities is incomplete in specific wavelength regions. Our O$_3$ opacity table begins at 1.4\,$\mu$m and does not include the broad visible ozone absorption feature spanning approximately 0.45-0.75\,$\mu$m \citep{Serdyuchenko2014}. Fourth, all planets are Earth twins with fixed mass, radius, and orbital impact parameter. Variations in surface gravity, scale height, cloud-top pressure, and atmospheric mixing can change the transmission signal substantially.

The detectability model introduces additional limitations. We assume photon-noise-limited $\sqrt{N}$ scaling from a reference stack, idealized treatment of telluric residuals, and a fixed duty-cycle prescription. Real HRCCS observations are affected by wavelength-correlated telluric residuals, time-variable stellar spectra, stellar spots and faculae, imperfect detrending, and the Rossiter-McLaughlin effect \citep{Birkby2018,BrogiLine2019,Palle2025}. This is particularly relevant for O$_2$, for which the planetary signal must be recovered in spectral regions containing strong terrestrial O$_2$ absorption. Moreover, the present detectability calculations use only the YJH channels of ANDES; consequently, the O$_2$ A-band at 0.76\,$\mu$m, as well as potentially useful optical CH$_4$ and H$_2$O features, do not contribute to the detectability maps. 

Nearby bright proxies also approach the soft-saturation limits. In addition, the photon budgets are calculated using the actual distances of the stellar proxies. The absolute observing times correspond to these specific, in several cases exceptionally nearby, systems and should not be interpreted as representative of a volume-limited FGKM population. The resulting requirements should be interpreted primarily as a comparative measure of molecular detectability. 

A further difference from \citet{Currie2023} is the treatment of exposure time. They set each simulated exposure equal to the full transit duration to minimize the contribution of read noise, whereas our exposure times are determined separately for each stellar proxy to avoid detector saturation while maintaining a high S/N. As a result, a single transit can contain many individual exposures in our simulations, and the conversion from the required number of frames to the number of transits depends on the assumed cadence and readout overhead. Six of the eleven stellar proxies considered here have exposure times shorter than 2\,s. Such short integrations may be difficult to realize in practice for full-frame observations. Differences in exposure-time prescription, together with the different noise treatments and detection metrics, also limit a direct comparison of our transit counts with those of \citet{Currie2023}.


\section{Conclusions}
\label{sec:conclusion}

We have investigated how the atmospheric evolution of an Earth-like planet would manifest in high-resolution transmission spectroscopy and how these signatures may be accessed with ELT/ANDES. We combine photochemical models spanning Archean, Proterozoic, and modern-Earth conditions with an FGKM host-star grid motivated by \citet{Rugheimer2018}, propagate the resulting atmospheric compositions through \texttt{petitRADTRANS}, and evaluate their high-resolution detectability using the \texttt{Ratri} framework. Because O$_2$ and CH$_4$ provide the clearest contrast between reduced and oxidized atmospheric states in our models, Figure \ref{fig:tracking_o2_ch4} summarizes how their joint detectability evolves across geological epochs and stellar hosts. Our main conclusions are summarized as follows:

\begin{enumerate}

\item The dominant high-resolution molecular signatures evolve substantially with atmospheric redox state. The CH$_4$-rich Archean models are characterized primarily by strong methane absorption, whereas O$_2$ becomes increasingly important following atmospheric oxygenation and reaches its strongest spectral contribution under modern-Earth conditions. The Proterozoic calculations demonstrate that this transition is gradual: atmospheres with very low O$_2$ abundances remain spectroscopically closer to anoxic states, while higher-O$_2$ cases develop increasingly prominent oxygen features.

\begin{figure}
\centering
\includegraphics[width=\columnwidth]{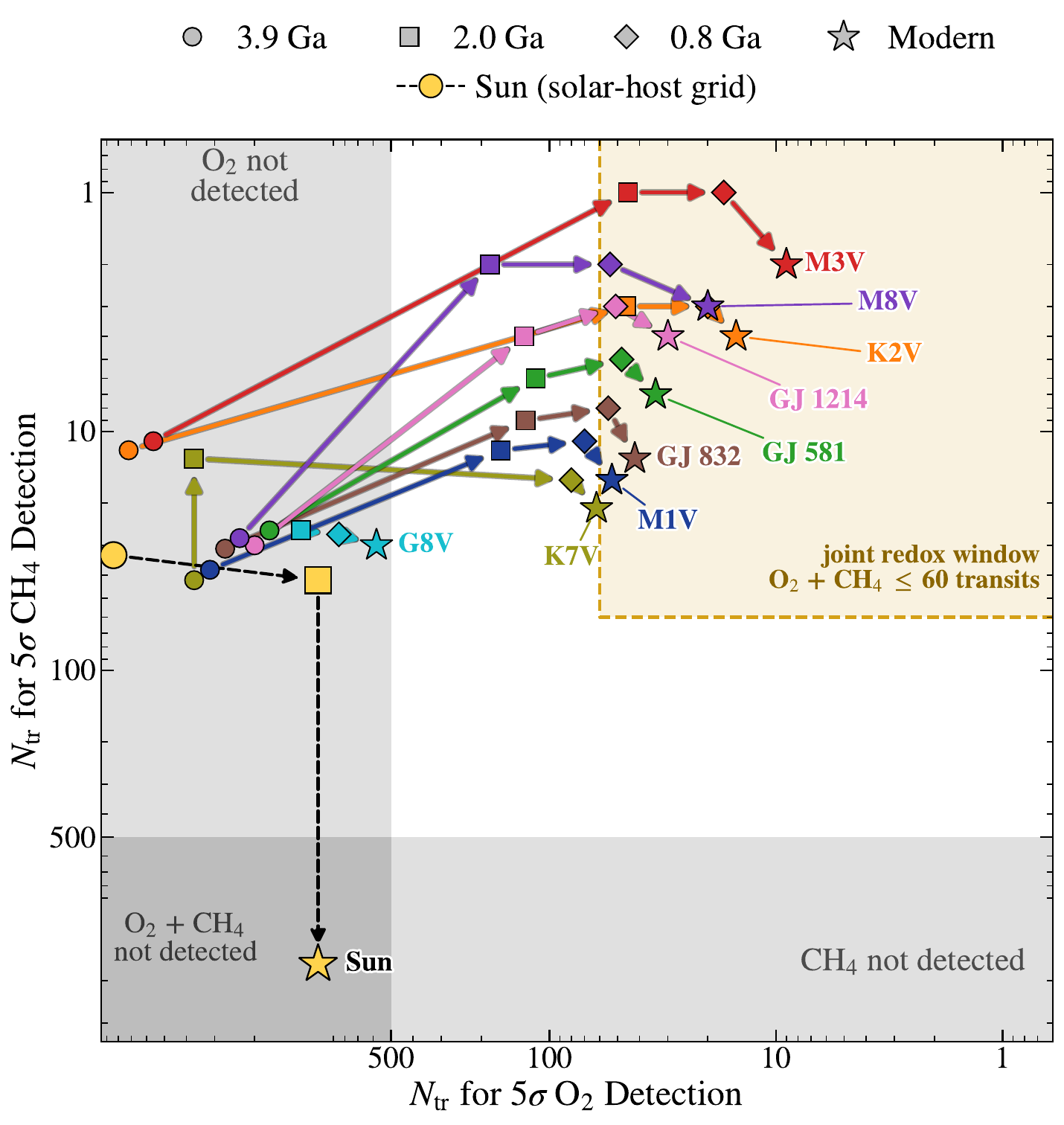}
\caption{Sumamry of 5$\sigma$ detection requirements with ELT/ANDES in the YJH band for O$_2$ and CH$_4$, following Figures \ref{fig:5sigma_modern_earth_detec} and \ref{fig:5sigma_detectability_diff_host}. The axes show the number of transits, $N_{\rm tr}$, required for 5$\sigma$ detections of O$_2$ and CH$_4$, with detectability increasing toward the upper right corner. Each colored track follows one host star through the four geological epochs of the FGKM grid (circle: 3.9\,Ga; square: 2.0\,Ga; diamond: 0.8\,Ga; star: modern), with arrows indicating the direction of evolution. Gray regions denote non-detections ($N_{\rm tr}>500$ or more than 200 equivalent observing nights), while the shaded yellow region marks cases where both O$_2$ and CH$_4$ are detectable within 60 transits. Hosts without any 5$\sigma$ O$_2$ or CH$_4$ detection are not shown. For the cooler proxies, the tracks move from O$_2$ non-detection at 3.9\,Ga into the joint O$_2$--CH$_4$ detection region after oxygenation, whereas O$_2$ remains outside the 5$\sigma$ observing limits for the solar-host cases. }
\label{fig:tracking_o2_ch4}
\end{figure}

\item CH$_4$ abundance alone does not provide a unique indicator of an Archean-like atmosphere. Although methane is particularly prominent in the methane-rich Archean models, substantial CH$_4$ can also persist in oxygenated atmospheres, especially around cooler host stars. The stellar UV radiation field modifies the photochemical lifetime of methane and therefore changes the relationship between geological state and observable CH$_4$ abundance. Consequently, a strong methane detection cannot by itself distinguish an anoxic early-Earth analogue from a later oxygenated atmosphere.

\item The host star influences both the atmospheric chemistry and the observable transmission signal. The stellar spectral energy distribution controls photolysis and thus affects the abundances of species such as CH$_4$, O$_3$, and N$_2$O, while the stellar radius directly determines the planet-to-star transmission contrast. These effects combine to produce substantial differences in molecular detectability across the FGKM host star grid. In our proxy sample, the shortest equivalent observing times generally occur for the M-dwarf proxies, although the nearby K2V case can also require relatively few transits for several molecules. The F-star cases are considerably more demanding because of their larger stellar radii and less favorable photon budgets.

\item The relative detectability of individual molecules depends strongly on both geological epoch and stellar host. O$_2$ does not reach the 5$\sigma$ criterion within the observing limits for any 3.9\,Ga atmosphere, but becomes gradually more detectable after oxygenation. In the modern M3V, K2V, and M8V cases, O$_2$ requires 9,15, and 20 transits, respectively, while CH$_4$ requires only 2,4, and 3 and CO$_2$ requires 5,6, and 13. H$_2$O is less accessible in the same cases, requiring 20,12, and 51 transits. Thus, the most favorable oxygenated cool-star systems can provide access to both oxidized and reduced species within feasible observing campaigns under our assumptions.

\item No single molecular detection uniquely identifies a geological state. The most informative diagnostic is the joint atmospheric context. In particular, the relative constraints on O$_2$ and CH$_4$ provide information on atmospheric redox state, while CO$_2$ and H$_2$O provide complementary constraints on the broader climatic and environmental conditions. A non-detection can also be informative only when the achieved abundance or spectral sensitivity is sufficiently constraining. Distinguishing atmospheric evolutionary states will need multi-molecule measurements.

\item The absolute observing-time estimates should be interpreted as optimistic lower limits. The present calculations assume clear atmospheres, idealized telluric correction, photon-noise-limited $\sqrt{N}$ stacking, one-dimensional atmospheric structures, and specific nearby stellar proxies. The detectability calculations are additionally restricted to the ANDES YJH channels and therefore exclude the strong O$_2$ A-band at $0.76\,\mu$m. These assumptions affect the absolute number of transits or observing hours required for detection.

\end{enumerate}

\begin{acknowledgments}
L.M. acknowledges funding support from the DAE through the NISER project RNI 4011. \\

Anthropic Claude Opus 5 was used to assist with figure generation, in accordance with the AAS guidelines on the use of large language models (\href{https://journals.aas.org/author-llm-guidelines}{https://journals.aas.org/author-llm-guidelines}). All scientific code and analyses were developed by the authors, who verified all results and take full responsibility for the intellectual integrity of the work presented.
\end{acknowledgments}

\begin{contribution}

D.D., S.D. and L.M. jointly developed the scientific framework and analysis plan. D.D. led the photochemical modeling, geological-grid analysis and transmission-spectrum calculations. \texttt{Ratri} ELT/ANDES simulation setup and subsequent high-resolution detectability analysis was performed by D.D. with inputs from S.D. The manuscript was written by D.D. and revised critically by S.D. and L.M. All authors contributed to the interpretation of the results and to the refinement of the final manuscript.
\end{contribution}


\software{astropy \citep{astropy:2013,astropy:2018,astropy:2022}, numpy \citep{harris2020array}, scipy \citep{virtanen2020scipy}, astroplan \citep{morris2018astroplan}, Photochem \citep{Wogan2025Photochem}, petitRADTRANS \citep{Molliere2019,Molliere2020}, Ratri \citep{dash2025detectability,dash2026probing}}, ESO SkyCalc \citep{noll2012atmospheric, jones2013advanced}

\newpage
\appendix

\section{Additional High-Resolution Transmission Spectra}
\label{app:spectral_atlases}

This appendix presents the detailed high-resolution transmission spectra underlying the trends discussed in Section \ref{sec:results}. The figures provide a more complete view of how the molecular features change across the Archean and Proterozoic atmospheric grids and illustrate the individual molecular contributions to the modern-Earth spectrum. 

\begin{figure*}
\centering
  \includegraphics[width=\textwidth]{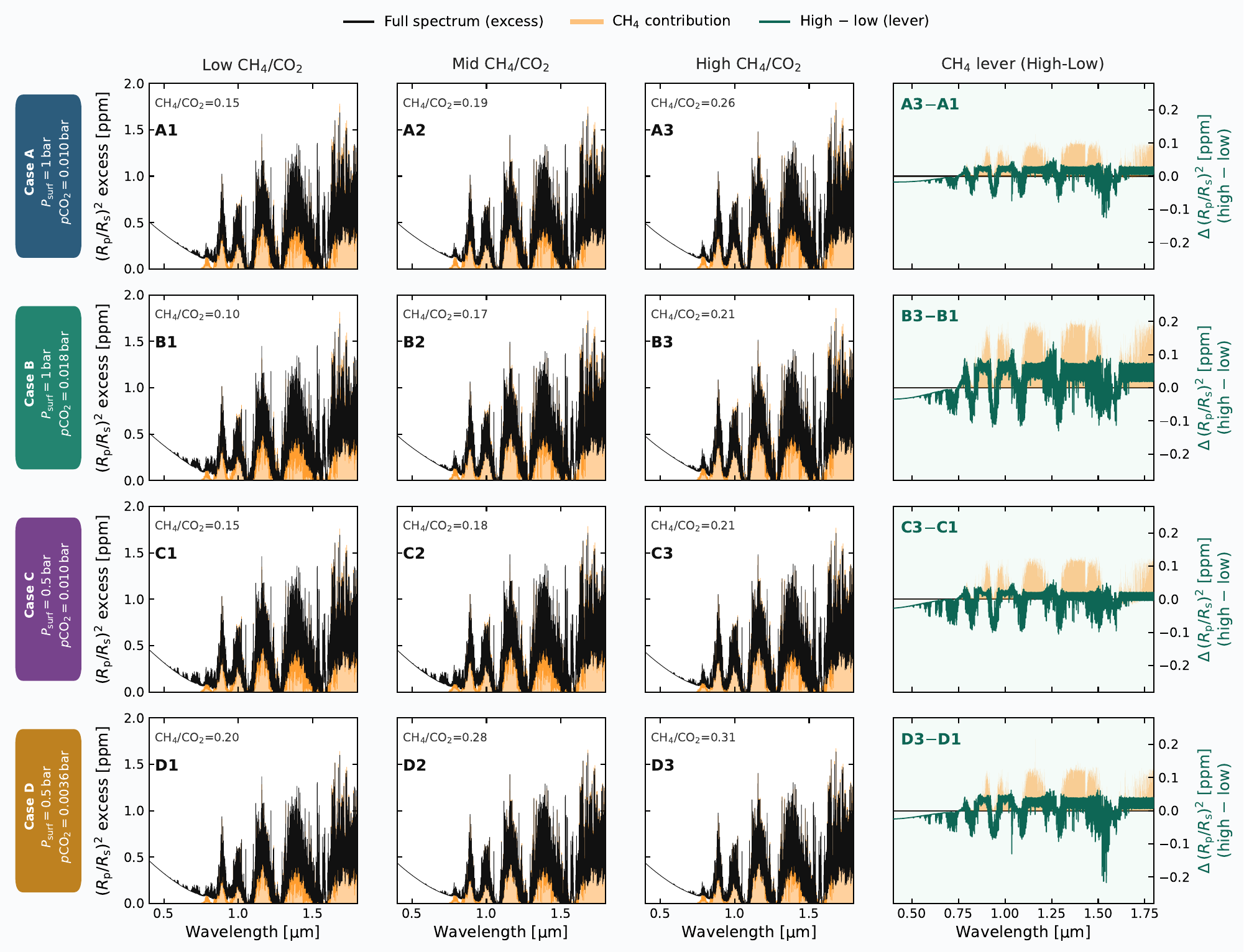}
  \caption{Archean spectral grid: rows are climate families A-D; the first three columns are rising surface CH$_4$/CO$_2$. Black curves show continuum-subtracted transit-depth excess; orange filled underlays show the CH$_4$ contribution in the spectra. The rightmost column shows the spectral change within each family between the highest and lowest CH$_4$/CO$_2$ cases: continuum-subtracted excess of the highest CH$_4$/CO$_2$ case minus that of the lowest (teal curve; e.g. column 3 $-$ column 1), with the orange underlay the corresponding high minus low CH$_4$ contribution. Positive (negative) values indicate stronger (weaker) features at higher CH$_4$/CO$_2$; the difference panels use a zoomed vertical scale.}
  \label{fig:archean_spectra}
\end{figure*}

\begin{figure*}
\centering
  \includegraphics[width=\textwidth]{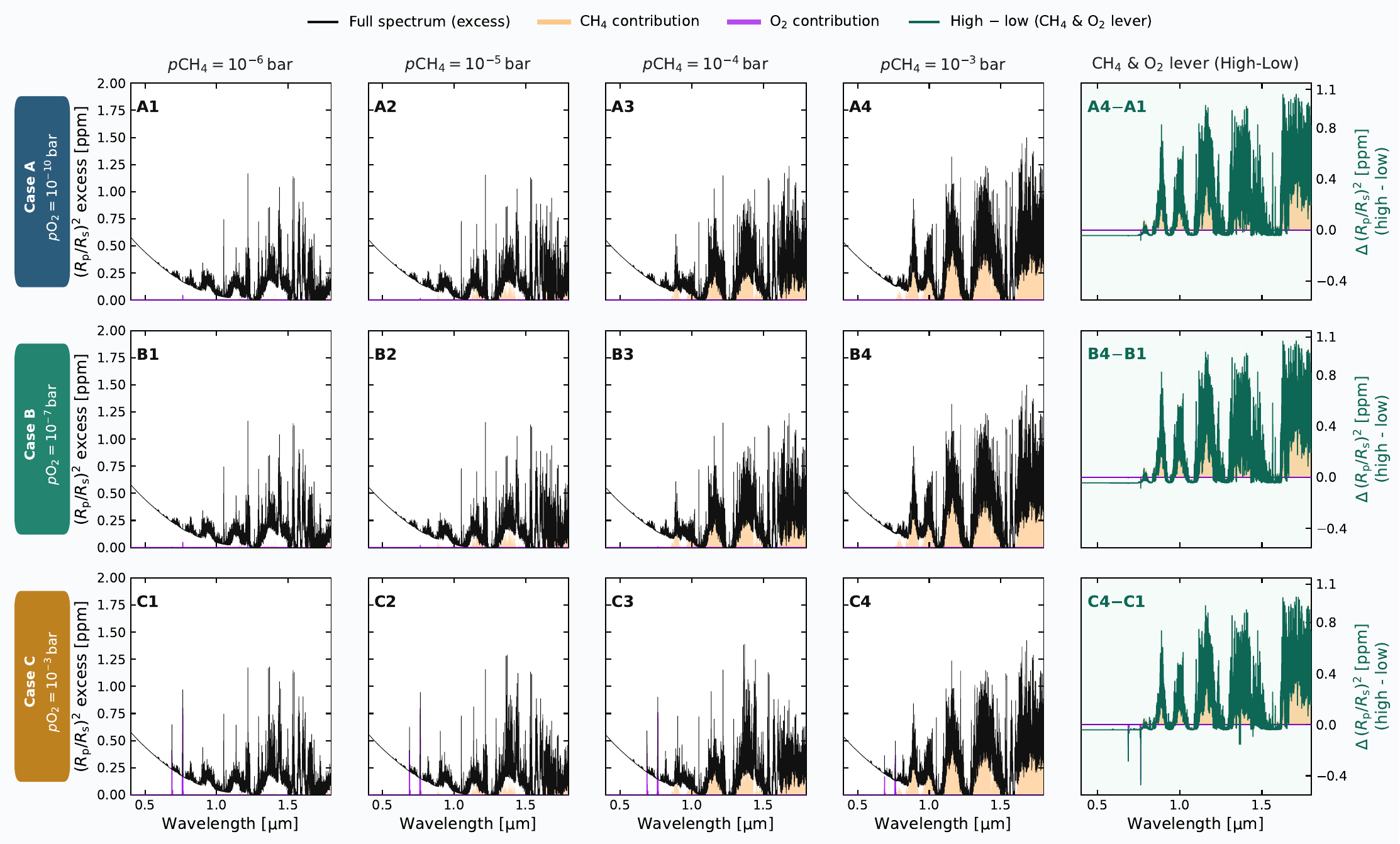}
  \caption{Proterozoic spectral grid: rows represent fixed surface $p\mathrm{O}_2$; the first four columns are rising surface $p\mathrm{CH}_4$. Black curves show continuum-subtracted transit-depth excess; orange and purple filled underlays show the CH$_4$ and O$_2$ contributions to the spectra, respectively. The rightmost column shows the spectral change within each O$_2$ row between the highest and lowest $p\mathrm{CH}_4$ cases: continuum-subtracted excess of the highest-$p\mathrm{CH}_4$ case minus that of the lowest (teal curve; e.g.\ column 4 $-$ column 1), with orange and purple underlays the corresponding high minus low CH$_4$ and O$_2$ contributions. Positive (negative) values indicate stronger (weaker) features at higher $p\mathrm{CH}_4$; the difference panels use a zoomed vertical scale.}
  \label{fig:proterozoic_spectra}
\end{figure*}

\begin{figure*}
\centering
  \includegraphics[width=\textwidth]{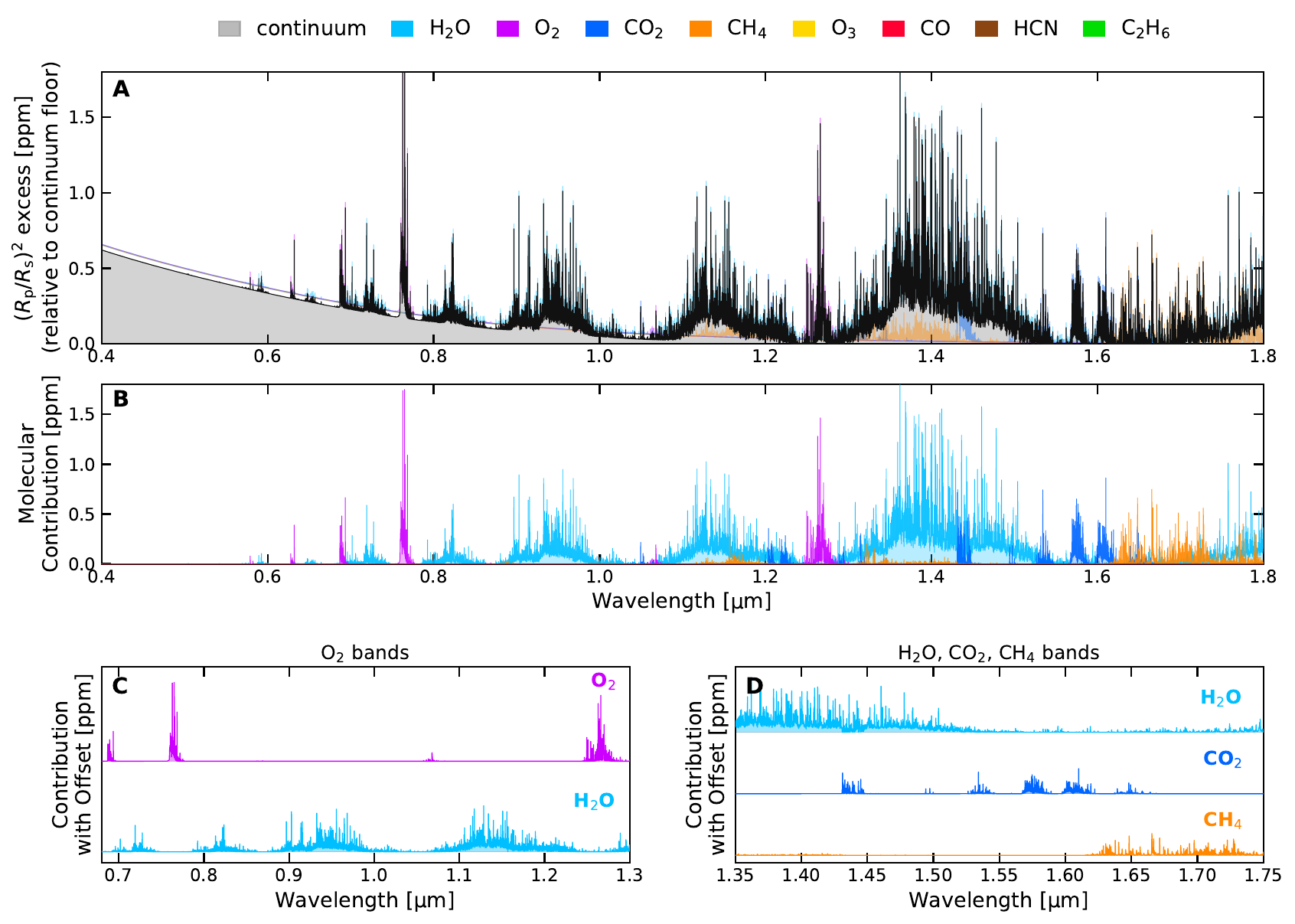}
  \caption{Modern Earth clear-atmosphere transmission spectrum at $R\sim10^{5}$ with  molecule-by-molecule attribution over $0.4$-$1.8\,\mu$m. In all panels, species are color-coded as in the legend. (\textbf{A}) Continuum-subtracted transit-depth excess $(R_{\mathrm{p}}/R_{\mathrm{s}})^{2}$ (black), measured relative to the 5th-percentile continuum floor (gray fill), with muted ``only-molecule'' traces of the active absorbers. (\textbf{B}) Molecular contributions $\Delta=({\rm full}-{\rm leave\text{-}one\text{-}out})$ in ppm for the same wavelength; O$_2$ and H$_2$O dominate selected optical-NIR intervals, while CO$_2$, CH$_4$, CO, HCN, C$_2$H$_6$, and O$_3$ provide weaker secondary features. (\textbf{C}) Offset contribution zooms over $0.68$-$1.30\,\mu$m isolating the O$_2$ B/A and $1.27\,\mu$m bands (upper) against overlapping H$_2$O (lower). (\textbf{D}) Offset contribution zooms over $1.35$-$1.75\,\mu$m through the near-infrared H$_2$O/CO$_2$/CH$_4$ forest (top to bottom). O$_3$ contributions appear only where the O$_3$ line list is available ($\lambda\gtrsim1.4\,\mu$m). Aerosols are not considered in the radiative transfer.}
  \label{fig:modern_spectra}
\end{figure*}

Figure \ref{fig:archean_spectra} shows the Archean spectral grid across the different climate families and CH$_4$/CO$_2$ boundary conditions. Figure \ref{fig:proterozoic_spectra} presents the corresponding Proterozoic grid, in which O$_2$ and CH$_4$ are varied independently. Figure \ref{fig:modern_spectra} provides a molecule-by-molecule decomposition of the modern-Earth spectrum and identifies the wavelength regions in which the principal absorbers contribute to the total transmission signal.

\clearpage

\section{Host-Star Dependence of the Molecular Spectra}
\label{app:host_spectra}

\begin{figure*}
\centering
  \includegraphics[width=0.9\textwidth]{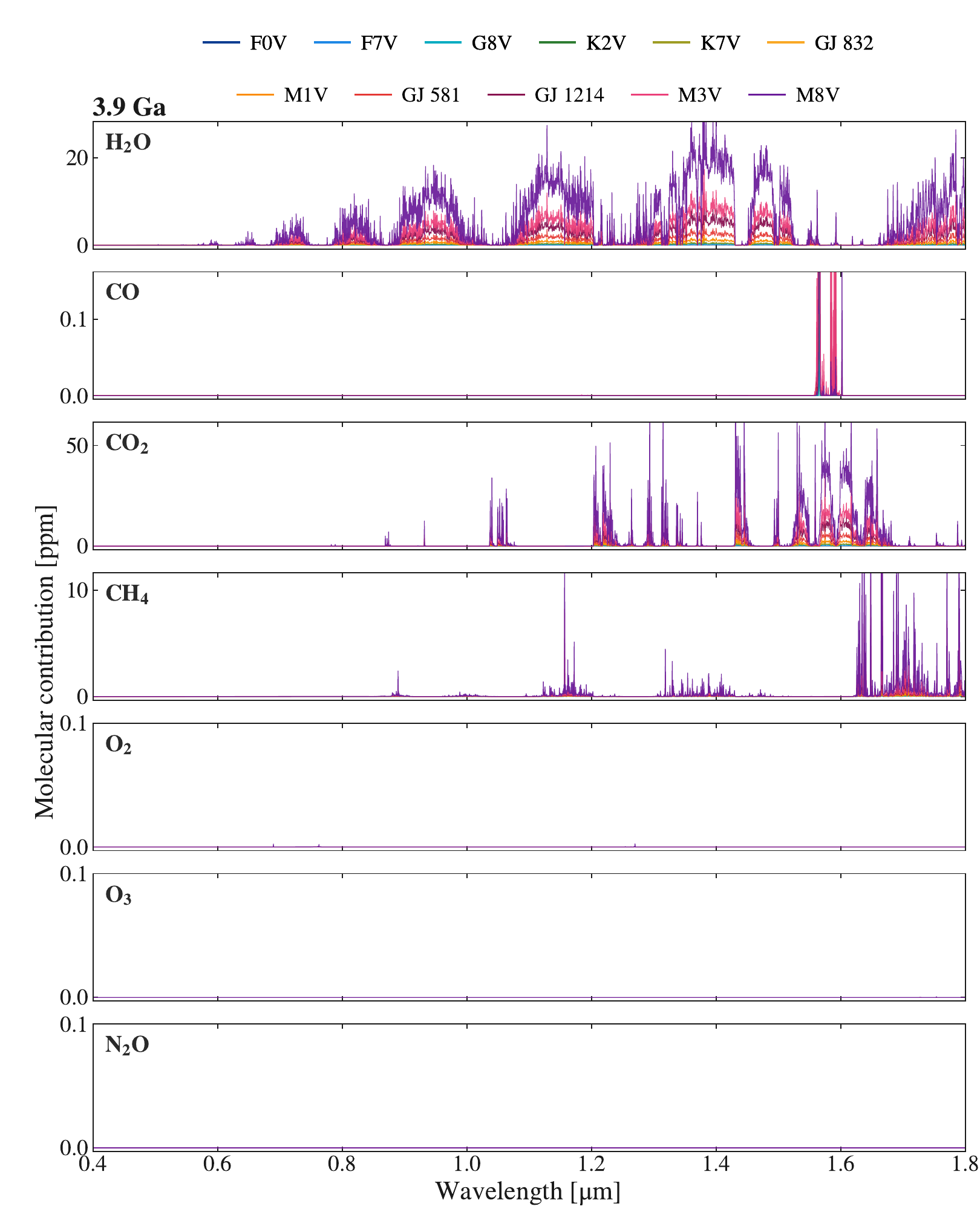}
  \caption{Molecule-isolated leave-one-out contributions for all eleven hosts over $0.4$-$1.8\,\mu$m for the anoxic 3.9\,Ga epoch. Hosts are colored by $T_{\mathrm{eff}}$ (cool M crimson to hot F blue). Features are CH$_4$- and CO$_2$-dominated; O$_2$ remains negligible.}
  \label{fig:appendix_3.9ga}
\end{figure*}

\begin{figure*}
\centering
  \includegraphics[width=0.9\textwidth]{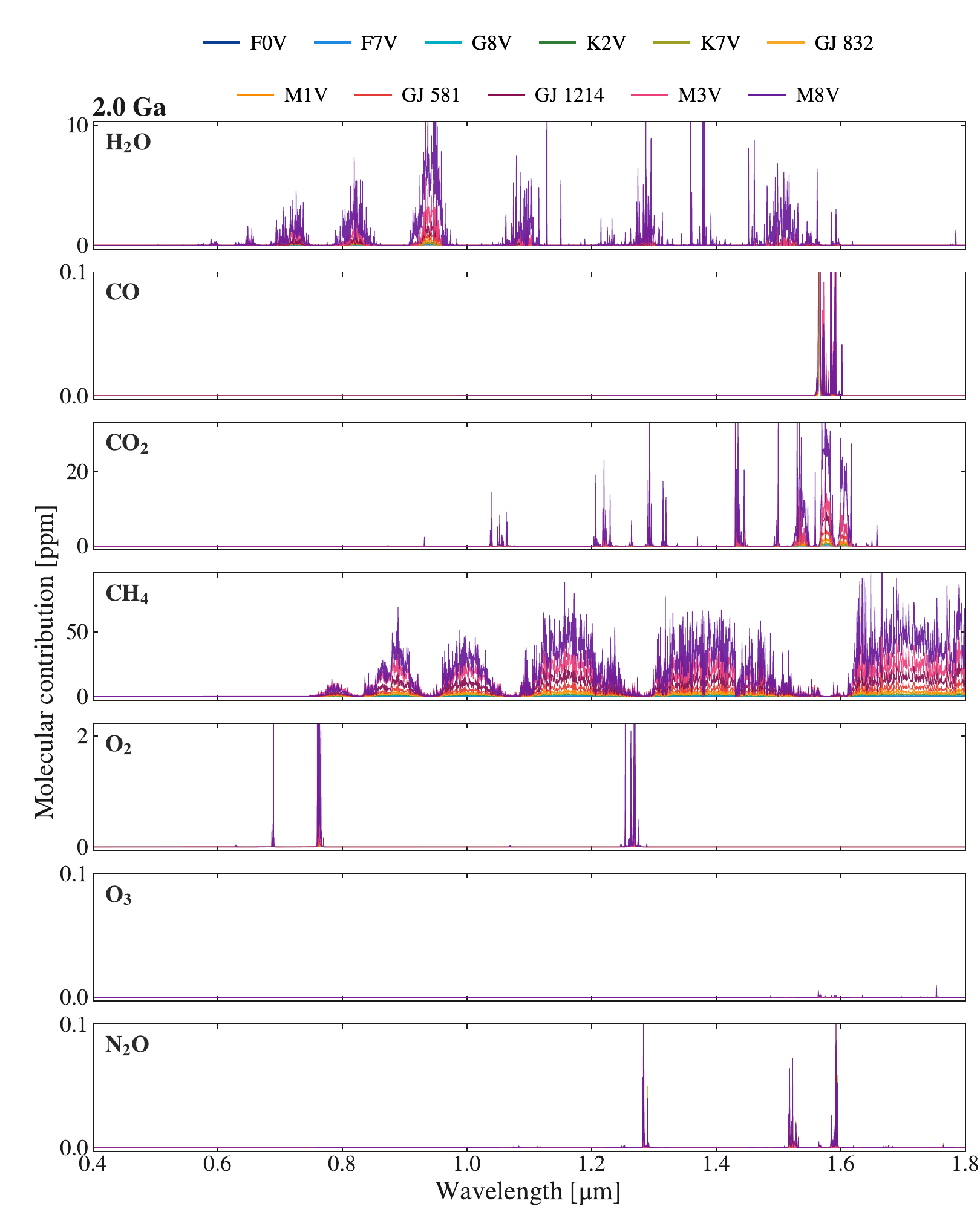}
  \caption{Same as Figure \ref{fig:appendix_3.9ga}, but for the 2.0\,Ga (post-GOE) epoch. O$_2$ contributions begin to appear while CH$_4$ remains strong, especially for cooler hosts.}
  \label{fig:appendix_2.0ga}
\end{figure*}

\begin{figure*}
\centering
  \includegraphics[width=0.9\textwidth]{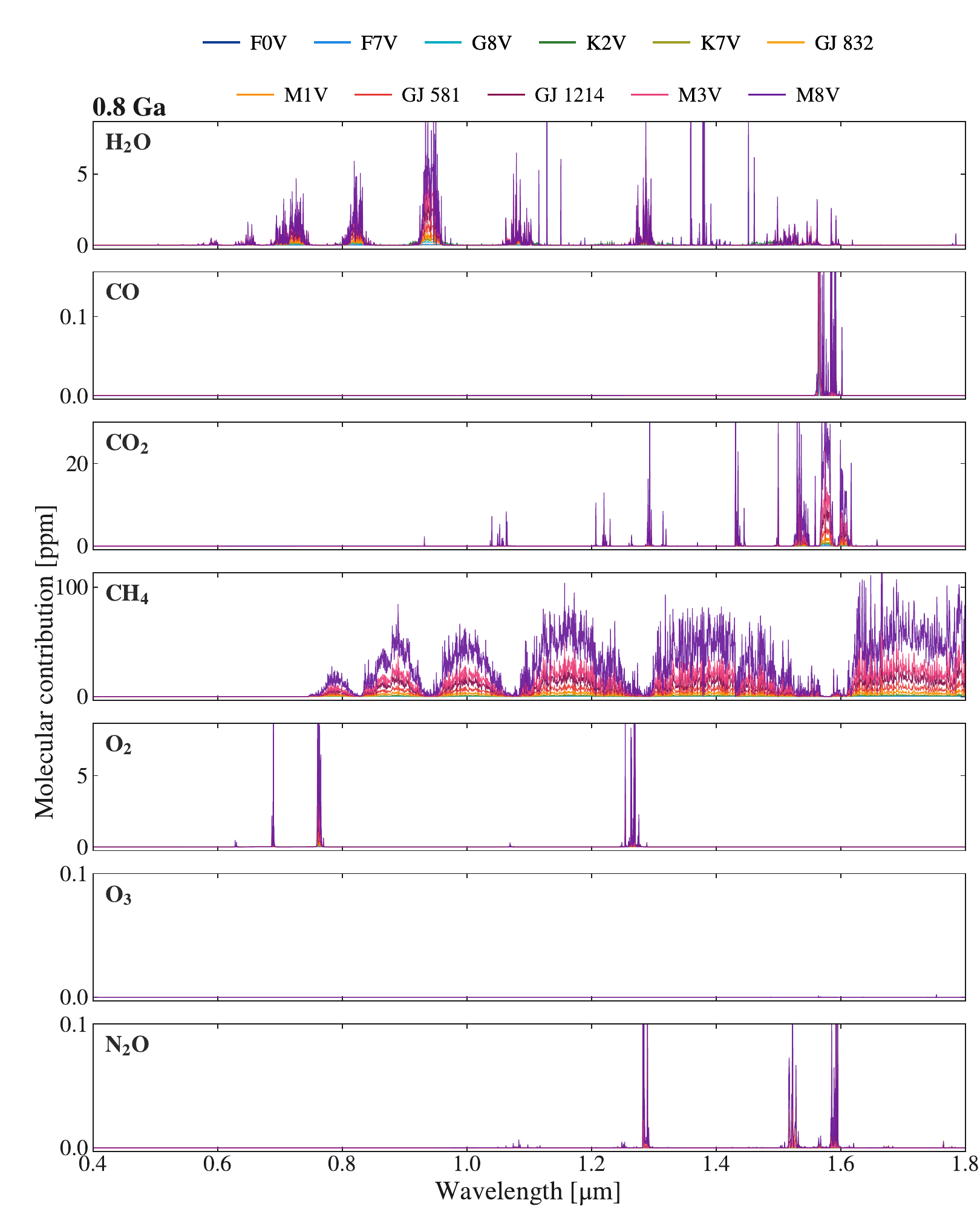}
  \caption{Same as Figure \ref{fig:appendix_3.9ga}, but for the 0.8\,Ga (Neoproterozoic) epoch at higher oxygen.}
  \label{fig:appendix_0.8ga}
\end{figure*}

\begin{figure*}
\centering
  \includegraphics[width=0.9\textwidth]{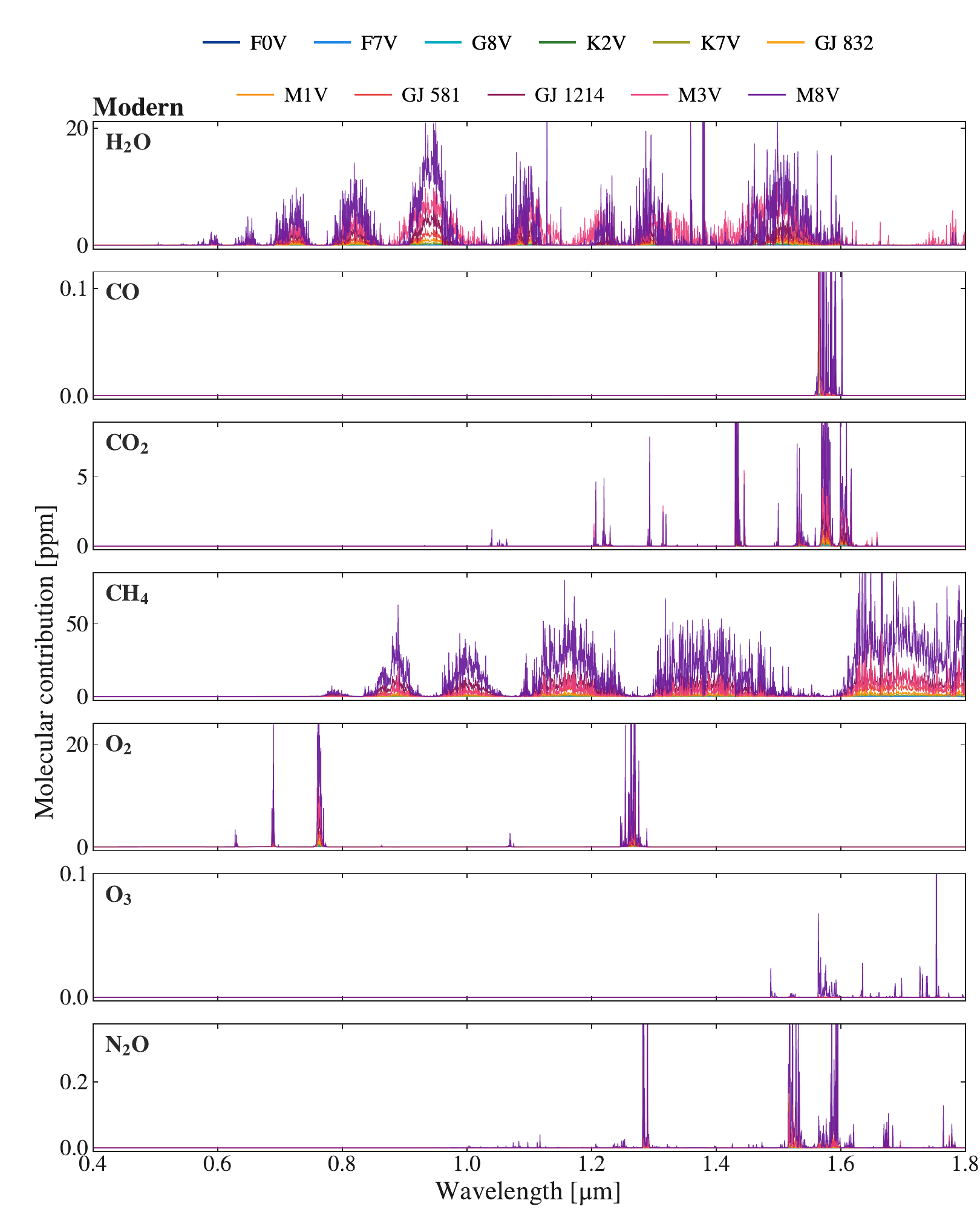}
  \caption{Same as Figure \ref{fig:appendix_3.9ga}, but for modern Earth boundary conditions. O$_2$ and H$_2$O dominate selected intervals; CH$_4$ is suppressed relative to earlier oxygenated epochs for many hosts.}
  \label{fig:appendix_modern}
\end{figure*}

Figures \ref{fig:appendix_3.9ga}-\ref{fig:appendix_modern} show the molecular contributions to the transmission spectra for the full stellar-proxy grid at each of the four geological epochs. These figures complement the host-star comparison presented in the main text by showing explicitly how the high-resolution molecular features change with stellar spectral type. The host-star dependence arises from both the different atmospheric compositions produced by the photochemical calculations and the changing stellar and planetary parameters used for the transmission spectra calculation. The figures are intended to illustrate the relative spectral behavior across the FGKM host star grid.

\clearpage

\section{ANDES Signal-to-Noise Ratios for the Stellar Proxy Grid}
\label{app:snr}

\begin{figure*}
\centering
  \includegraphics[width=\textwidth]{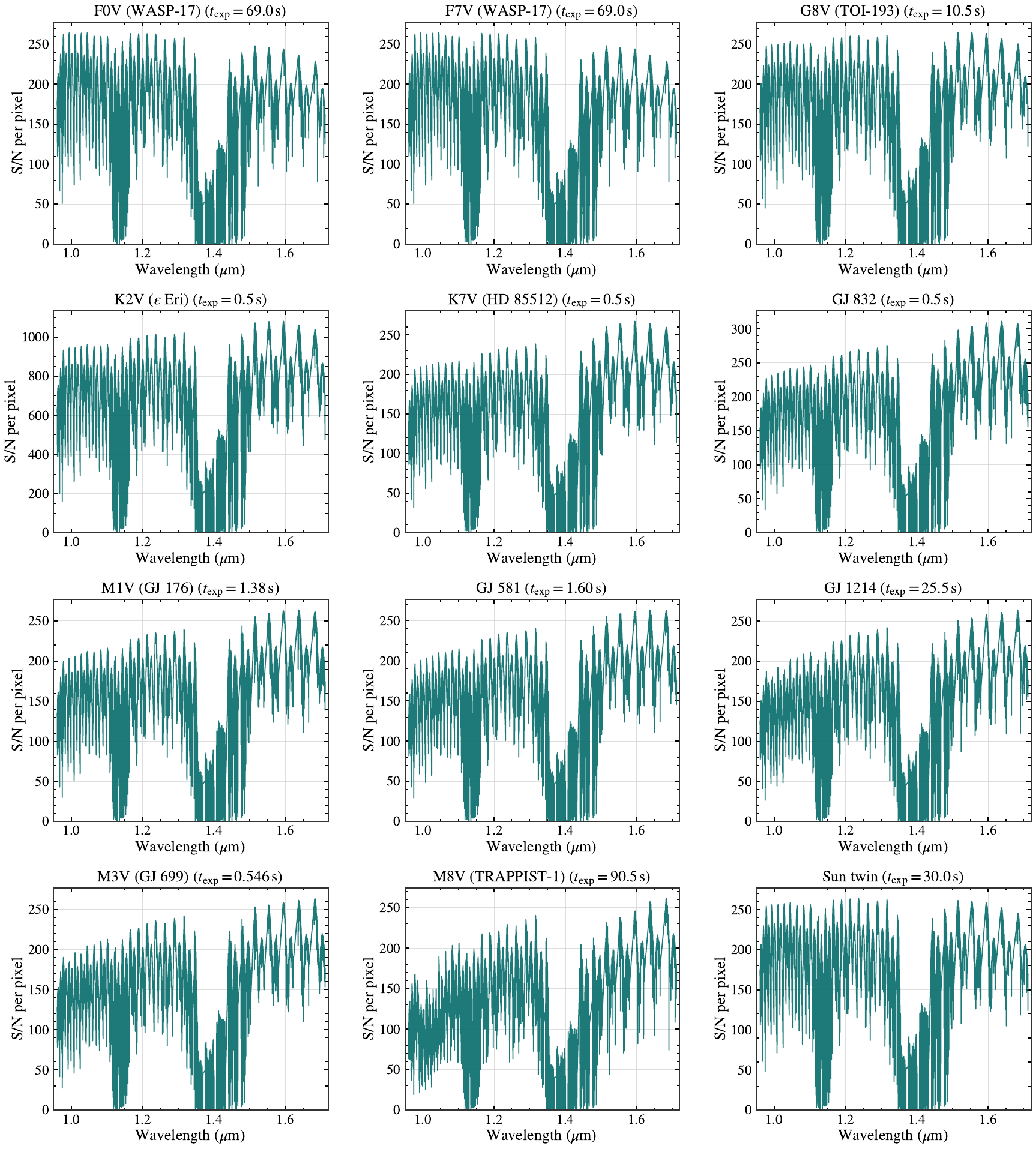}
  \caption{Continuum signal-to-noise ratio per pixel as a function of wavelength for the ANDES YJH mode at each host's recommended exposure time ($t_{\mathrm{exp}}$). Exposures are auto-scaled so the stellar continuum reaches an assumed unsaturated level of $\sim7\times10^{4}\,e^{-}$. Most hosts sit in that regime; the K2V proxy ($\varepsilon$ Eri) remains soft-saturated even at the $0.5\,\mathrm{s}$ floor, and GJ\,832 at the same floor lies near soft saturation. Toward cooler hosts, the relative $S/N$ shifts toward longer wavelengths within the YJH range, with the H band becoming increasingly favorable relative to the Y and J bands.}
  \label{fig:snr_host}
\end{figure*}

Figure \ref{fig:snr_host} shows the wavelength-dependent continuum signal-to-noise ratio adopted in the \texttt{Ratri} simulations for each stellar proxy. The exposure times are adjusted to approach the assumed continuum count level while avoiding saturation where possible. This figure provides the instrumental context for the host-dependent detectability differences discussed in Section \ref{sec:results}. For the brightest nearby systems, the minimum exposure-time constraint limits this scaling. In particular, the K2V proxy $\varepsilon$ Eri remains above the nominal unsaturated count level at the 0.5\,s exposure-time floor, while GJ 832 lies close to this regime.

\clearpage
\bibliography{sample701}{}
\bibliographystyle{aasjournalv7}



\end{document}